# PORTABLE RECORDING SYSTEM FOR SPHERICAL THERMOGRAPHY AND ITS APPLICATION TO LONGWAVE MEAN RADIANT TEMPERATURE ESTIMATION

Takashi Asawa[a*], Haruki Oshio[a], Kazuki Tanaka[b]

[a] School of Environment and Society, Tokyo Institute of Technology, 4259 Nagatsuta-cho, Midori-ku, Yokohama, Kanagawa 226-8502, Japan

[b] Tokyu Construction Co., Ltd., 1-16-14, Shibuya, Shibuya-ku, Tokyo 150-8340, Japan

[*]Corresponding author. Tel.: +81 459245510, Fax: +81 459245553. E-mail address: asawa.t.aa@m.titech.ac.jp

Postal address: School of Environment and Society, Tokyo Institute of Technology, 4259-G5-2 Nagatsuta-cho, Midori-ku, Yokohama, Kanagawa 226-8502, Japan

**Abstract**

Mean radiant temperature (MRT) is the primary metric of radiant heat exchange between a human body and the environment, and it dominates human thermal comfort and heat stress. This study develops a new portable recording system for spherical thermography utilizing only commercial devices (an infrared thermal imaging camera and a portable rotation platform) and image processing for panorama synthesis. We use the system to estimate the MRT in a longwave radiation environment. A spherical thermal image is generated by synthesizing 24 source images on the basis of feature point identification. The longwave MRT and plane radiant temperature can be estimated from the spherical thermal image using image projection methods, including the orthographic projection and Lambert cylindrical projection. To validate the developed system, measurements were made using the system in outdoor and indoor environments with various radiant temperature distributions, including sunny built spaces and a tree-shaded space, and the results were compared with those obtained using pyrgeometers. The difference in longwave MRT between the estimation by the developed system with orthographic projection of spherical thermal images and the measurement by pyrgeometers for six directions was within 1 °C in most cases and 1.6 °C at maximum. The results show that the developed system has sufficient accuracy for longwave MRT estimation while evaluating the radiant temperature distribution and radiant asymmetry of spaces.

## 1. Introduction

Human bodies exchange energy with the environment by conduction, convection, evaporation (i.e., sweating), and radiation. These energy exchanges are the main factors determining the thermal comfort and heat-related stress of the human body in both cold and hot environments. Many studies have demonstrated that half of thermal comfort is driven by radiant heat exchange [1], and the importance of convection relative to radiation is approximately 1.0, with a range of 0.71 to 1.4 [2]. These studies indicate the importance of radiant heat exchange for understanding human thermal comfort. Radiant heating/cooling systems are effective for stably maintaining the thermal comfort of occupants of indoor spaces [3,4], because the surface temperatures of heating/cooling panels and view factors from the human body to these panels are the primary factors affecting radiant heat exchange. In these systems, radiant heat exchange occurs by longwave radiation (thermal infrared radiation). In outdoor spaces, radiant heat transfer occurs by shortwave radiation (i.e., solar radiation) and longwave radiation. Shortwave radiation dominates human thermal comfort [5] and the use of outdoor spaces [6,7] under sunny conditions. In addition, longwave radiation from the ground and building walls, which have higher temperatures because they receive shortwave radiation, causes significant radiant asymmetry and affects the radiant heat exchange between a human body and its environment [2]. During nighttime in hot climates, atmospheric radiative cooling enhances human thermal comfort under a clear sky [8], because the downward atmospheric radiation (longwave radiation) from the sky is generally lower than the longwave radiation emitted from the human body owing to the existence of an atmospheric window (8–14 μm).

Mean radiant temperature (MRT) is the primary metric of radiant heat exchange between a human body and the environment. The MRT is defined as the "uniform surface temperature of an enclosure in which an occupant would exchange the same amount of radiant heat as in the actual nonuniform enclosure" [9]. Radiant heat exchange between the human body and its surroundings involving both shortwave and longwave radiation can be explained in terms of the MRT. Therefore, the MRT is widely used to evaluate human thermal comfort [10–14] and heat stress [15–18] in both indoor and outdoor spaces. There are many challenges in estimating the MRT with different measurement and estimation methods. The details of various methods and instruments, including their accuracy requirements, are reported in ISO Standard 7726 [19]. However, difficulties arise when researchers try to accurately estimate the MRT because of the complexity of urban and architectural geometries, asymmetry and nonuniform radiant temperature within spaces, and complexity of human body shapes [2]. In addition, each measurement and estimation method has specific targets and limitations.

The black globe thermometer is the most common method of measuring the MRT owing to its simplicity [20,21]. The equilibrium temperature of the globe depends on convective and radiative heat transfer through the globe thermometer [21]; therefore, these effects should be separated by measuring air temperature and wind speed simultaneously [22]. The globe thermometer method has fundamental errors related to wind speed [22], response time, diameter, and materials [21]. The most accurate way of determining the outdoor MRT is to measure the three-dimensional shortwave and longwave radiation fields and the angles between a person and the surrounding surfaces [23,24]. Typically, three net radiometers, each measuring the four radiation components separately, i.e., the shortwave (pyranometer) and longwave (pyrgeometer) incoming and outgoing radiation fluxes, are mounted on a stand. A total of 12 sensors are needed; therefore, this method is accurate but more expensive and less portable than a globe thermometer. The orthogonal instrument setup may cause instrumental error at high angles of incidence [23]. An attempt has also been made to estimate the MRT distribution in a room with a simple structure consisting of six surfaces by estimating the temperature of each surface using a radiometer [25]. In another study, MRT values obtained using a black globe thermometer and net radiometers have been compared [26].

In recent years, several methods of estimating the longwave MRT ($MRT_l$) using infrared thermography have been developed. An infrared thermal imaging camera (IR camera) can remotely observe the radiant temperature of an object even if it is not at the exact location of the object. A spherical thermal image ($4\pi$ image) can be obtained by rotating the camera horizontally and vertically (panning/tilting) using servos, and the $MRT_l$ can be calculated from the image. Compared to the black globe thermometer and net radiometer, the thermographic method can only evaluate longwave radiation; its great advantage is that it can visualize radiant temperature distribution and evaluate the impact of each surface on MRT. This is quite important for designing the thermal environment of urban/built spaces, as mentioned above. The first trial was performed by Asano et al. [27]. They developed a spherical thermography system with an IR camera and a custom-made platform that was rotated by an electric motor, and they calculated the $MRT_l$ in an outdoor space. Paz y Miño et al. [28] applied a similar method to obtain a spherical thermal image and calculate the MRT. They calculated the MRT of outdoor environments using both longwave and shortwave radiation on the basis of infrared thermography (8–14 μm) and high-dynamic-range photography (380–780 nm). Lee and Jo [29] also proposed a scanning system with an IR camera. They estimated MRT in indoor space using the scanning system and shortwave radiation estimated using the global irradiance measured outdoor and the three-dimensional geometry of the room [30,31]. Revel et al. [27] developed a system using a scanning linear array of thermopiles installed on the ceiling. To obtain the spatial distribution of the MRT, the radiant temperature distribution measured by the sensor and the three-dimensional geometry of the target space were combined [27,29,32]. Systems that simultaneously measure the radiant temperature and three-dimensional geometry using an infrared temperature sensor and a laser range sensor, such as the Spherical Motion Average Radiant Temperature sensor developed by Guo et al. [33], may simplify the process of integrating the radiant temperature and three-dimensional model, although the readability of the radiant temperature distribution may be inferior to that of thermal images. If the target space is small and has a simple spatial geometry, measurements at one or several points are sufficient to capture the radiant temperature of all surfaces. Measurements at multiple points are required for complex spatial geometries, such as actual urban spaces. However, existing methods require heavy devices and professional/expert systems that only developers and some researchers can use.

Therefore, this study develops a new portable recording system for spherical thermography using only commercial devices and image processing for panorama synthesis. We used the system for the practical estimation of the $MRT_l$, and the calculated $MRT_l$s were validated using pyrgeometers.

## 2. Development of the system

We developed a system for obtaining spherical thermal images by combining a small IR camera and rotation platforms. Figure 1 shows the components and assembled system. The system is very compact and lightweight. The following subsections describe the instruments, control of pan-tilt scanning, image synthesis to obtain a spherical image, and calculation of the $MRT_l$.

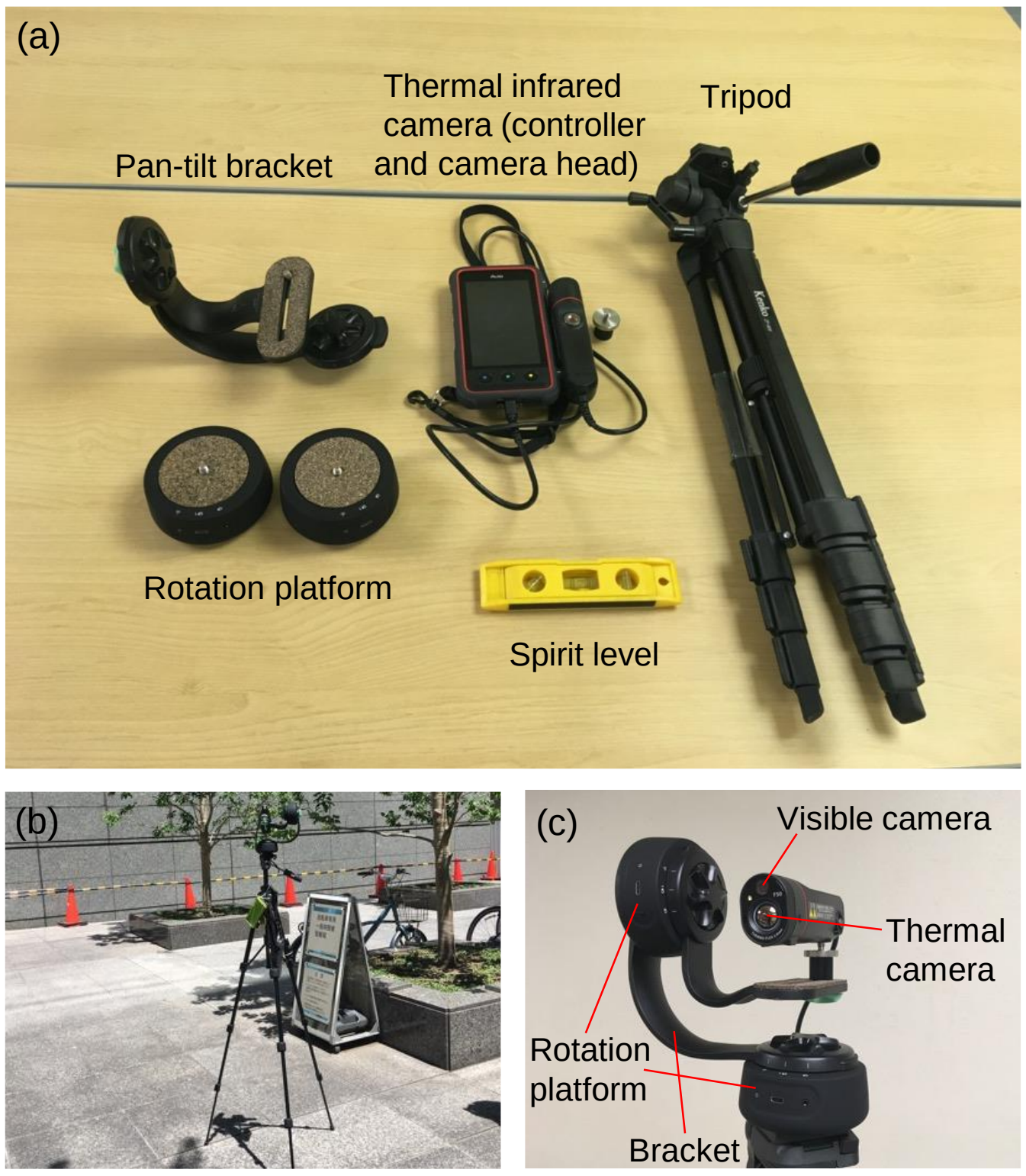

Fig. 1. Photographs of the developed system: (a) instruments comprising the system; (b) appearance of the system; (c) magnified view of camera part.

### *2.1 Instruments*

The specifications of the instruments are shown in Table 1. In the IR camera, the camera head was separated from the controller and connected by a cable. Because the camera has a wide field of view (70° by 70°), we can generate a spherical image from a small number of source images. The measurement wavelength range of the camera is 8–14 μm, which is similar to that of other general IR cameras. The camera records thermal and visible images simultaneously. The battery is built into the camera and has a lifetime of approximately 4 h. The weight of the camera is only 500 g (100 g for the camera head and 400 g for the controller). The rotation platform was adopted because it is small and commercially available and can operate as required according to the field of view of the camera. The rotation angle and rotation speed can be configured using an app compatible with iOS and Android. The battery is built into the device and has a lifetime of 8–14 h. To allow rotation around two axes, we combined two rotation platforms using a pan-tilt bracket (SY0003-0001; Vitec Imaging Solutions SPA, Cassola, Italy) [Fig. 1(c)]. The IR camera was mounted on the platform for tilt rotation, and the platform for pan rotation was mounted on a tripod. The total weight is 2.7 kg; a user can carry the system by one hand by holding the tripod. The angle of the camera head was adjusted using a spirit level so that the camera head was level when the tilt angle was 0°.

Table 1 Specifications of the developed system.

| | | |
|---|---|---|
| Entire developed system | Number of source images | 24 |
| | Recording range | 360° (Pan) × 160° (Tilt) |
| | Total weight | 2.7 kg |
| | Recording time | 2 min |
| Thermal infrared camera: Thermo FLEX F50 (Nippon Avionics Co., Ltd., Yokohama, Japan) | Field of view | 70° × 70° |
| | Measuring temperature range | −20–350 °C |
| | Spatial resolution | 5.3 mrad |
| | Accuracy | ±2 °C or 2% |
| | Sensitivity (temperature resolution) | 0.05 °C at 30 °C |
| | Spectral range | 8–14 μm |
| | Battery | Built-in (lasts 4 h) |
| | Recording pixels | 240 × 240 (Thermal) |
| | | 720 × 720 (Visible) |
| Rotation platform: Genie Mini II (Vitec Imaging Solutions Spa, Cassola, Italy) | Maximum rotation speed | 11° $s^{-1}$ |
| | Control | Mobile app |
| | Battery | Built-in (lasts 8–14 h) |

### *2.2 Pan-tilt scanning*

Thermal images were recorded at 45° intervals of both the pan and tilt angles (Fig. 2). Tilt angles from −90° to −70° were excluded to avoid having the tripod in the image. A total of 24 source images were recorded and used to generate a spherical image. The field of view of the camera is 70°; therefore, neighboring images overlapped by 25° [Fig. 2(a)]. This overlap was used to extract feature points and to synthesize images (Section 2.3). We configured the app so that the platform behaved as shown in Fig. 2(b). The camera first rotated in the pan direction (0°–315°) with a fixed tilt angle. Then, the camera rotated in the tilt direction by 45°. The pan rotation speed was set to 45° per 5 s, considering the maximum rotation speed of the platform of 11° $s^{-1}$. In line with this, the recording interval of the IR camera was set at 5 s. Rotation and recording were performed independently according to each setting. Therefore, rotation and recording started simultaneously (the start buttons of the app and camera were pressed at the same time) to synchronize them as much as possible. Different settings were used for the tilt rotation and the subsequent pan rotation: 45° per 4.5 s and 45° per 5.5 s, respectively. These settings were to ensure that source images 9 and 17 were recorded after the tilt rotation was complete, even if there was a slight difference (within 0.3 s) between the start times of the camera and platform. We considered that the source images 9 and 17 would line up with other images in the same tilt angle without misalignment by these settings, which would allow better image synthesis. It was confirmed that the rotation speed was slow enough for the camera recording, given that there was no motion blur in the recorded images. The time required to record 24 source images was approximately 2 min. Fig. 2(c) shows the app during configuration.

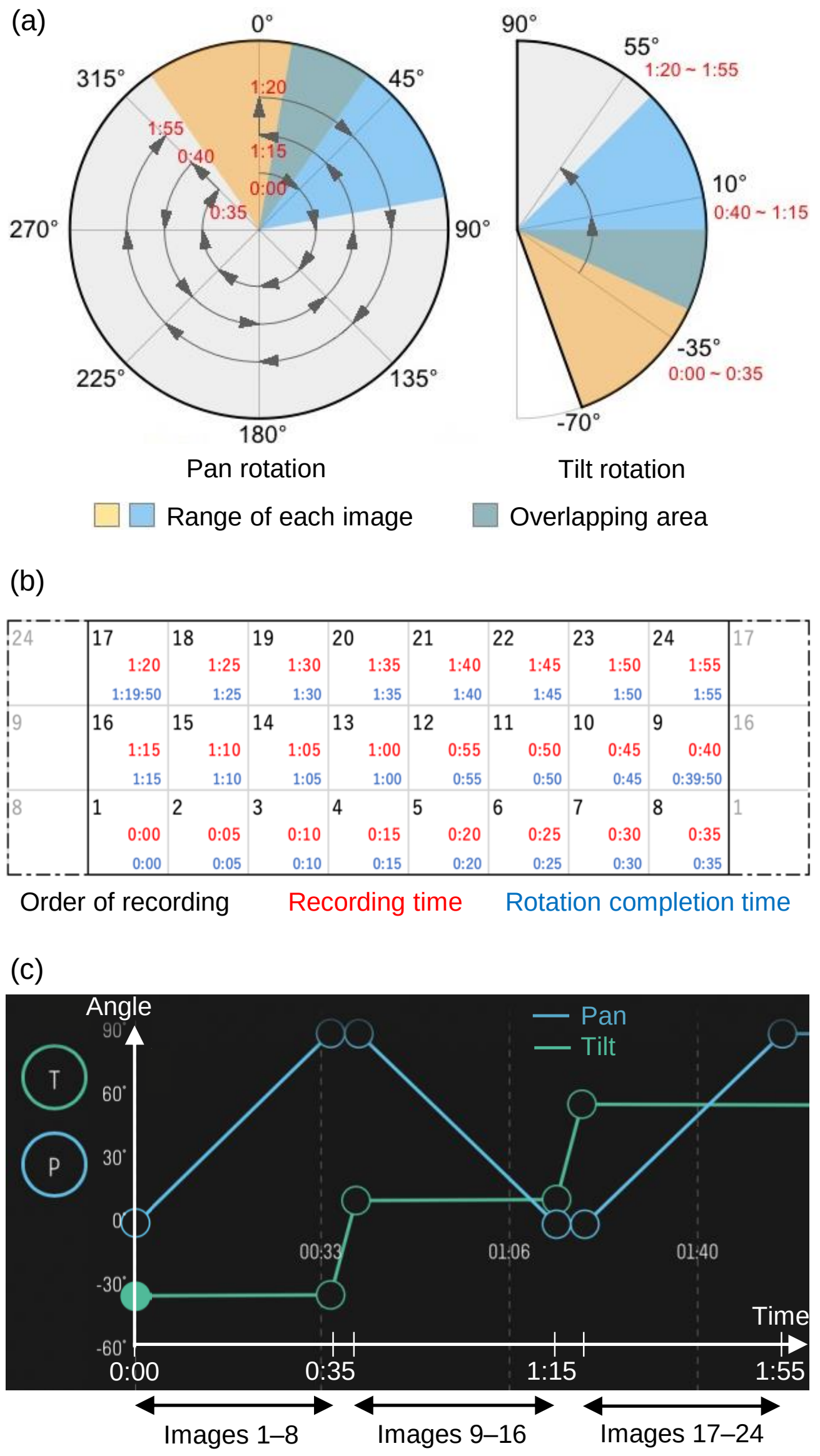

Fig. 2. Schematic of pan-tilt scanning: (a) rotations in pan and tilt directions and range of each image; (b) order and time of the recording of source images; (c) screenshot of app during setup (vertical axis indicates tilt angle).

### *2.3 Synthesis of images*

The recorded images were exported from the camera to the computer to generate a spherical image. It was difficult to accurately determine the recording angle of each image because the rotation of the platform and recording by the camera were not perfectly synchronized. Therefore, the panorama was synthesized by image matching using extracted feature points in the thermal images, rather than geometrically synthesized using information on the recording angle. Image stitching software (PTGui; New House Internet Services B.V., Rotterdam, Netherlands) was used for the synthesis. In the software, feature points can be identified automatically or manually on the basis of radiant temperature distributions. In our system, feature points were first identified automatically, and then additional points were identified manually if the number of points was insufficient for accurate synthesis. The boundaries between surfaces with different temperatures were used as feature points in manual processing. The software can create a spherical image using only the feature points, but to improve the quality of the synthesis, we input the recording angle of each image as supplemental information. The input recording angle was the angle assumed from the settings [e.g., the pan and tilt angles are 135° and −35°, respectively, for the fourth image of 24, as shown in Fig. 2(a),(b)] and was not the actual one; however, this information can facilitate the identification of feature points by the software. Figure 3 shows the feature identification process in the software and an example of the generated spherical image. In Fig. 3(b), the spherical image is presented in the equirectangular projection. By changing the projection method, the $MRT_l$ of a human body approximated by a sphere or cuboid can be calculated.

(a)

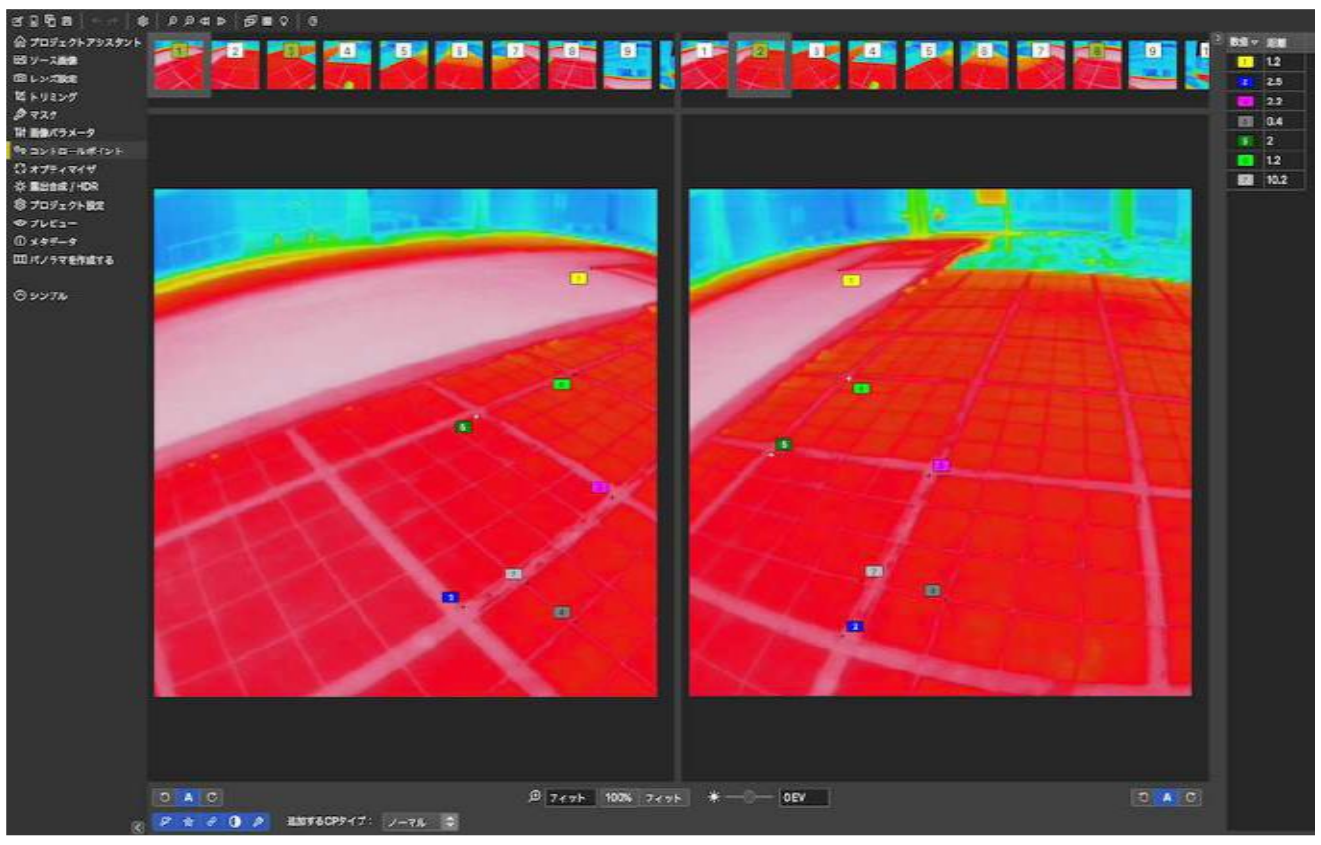

(b)

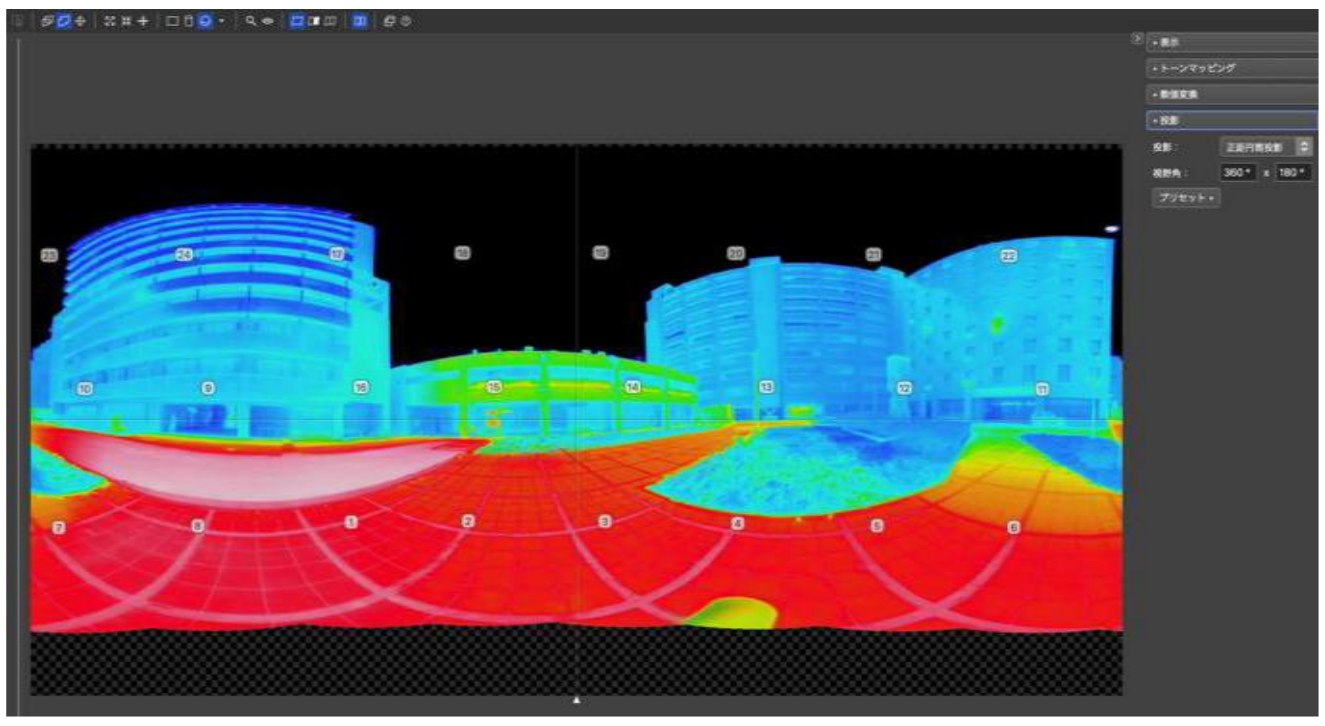

Fig. 3. Image synthesis using stitching software: (a) identification of feature points; (b) synthesized spherical image from 24 source images.

### *2.4 Post-processing*

#### *2.4.1 Retrieval of radiant temperature value from synthesized images*

The synthesized spherical thermal image had only color information but did not retain radiant temperature values in the bit number. To calculate the $MRT_l$, the radiant temperature was retrieved from the color information. For a color in the synthesized image that was included in the color scale used for the 24 source images, the radiant temperature value was determined to be the temperature corresponding to the color [for the colors assigned to temperatures from $T_1$ to $T_2$, the radiant temperature was given by $(T_1 + T_2)/2$]. Synthetic colors were generated where source images overlapped. For each synthetic color, we determined which color on the color scale was closest to the synthetic color by obtaining the RGB value minimizing the residual sum of squares, $S^2$:

$$S^2 = (r - r')^2 + (g - g')^2 + (b - b')^2, \qquad (1)$$

where $r$, $g$, and $b$ are the RGB values to be determined (for any one of the colors in the color scale), and $r'$, $g'$, and $b'$ are the RGB values of the synthetic color.

*2.4.2 Correction for blind areas*

Images were not obtained at tilt angles of −90° to −70° to avoid having the tripod in the image. The mean value of the radiant temperature in pixels adjacent to this blind area [which appears as an area with white dots on a black background in Fig. 3(b)] was assigned to the pixels of the blind area. When the $MRT_l$ is calculated assuming it is that of a human body, this area can be removed because it is within the foot area. For the area around the zenith direction (tilt angle approximately 90°), a small area was found to have no valid temperature value as a result of image synthesis in some cases. The same correction was applied to that area.

*2.4.3 Correction for sky radiant temperature*

The measurement wavelength range of the camera is 8–14 μm; in this range, the atmospheric transmittance is high. Therefore, the $MRT_l$ is underestimated if the measured radiant temperature is used as obtained. The temperature was corrected using a method similar to that of Asano et al. [27]. The radiant emittance of a blackbody at the measured temperature was calculated for 8–14 μm using Planck's law. The radiant emittance of a blackbody at the air temperature was calculated for 3–8 μm and wavelengths longer than 14 μm. The atmospheric emittances in these wavelength ranges are close to 1; therefore, most radiation in these regions is emitted from the adjacent atmosphere. These radiant emittance values were summed, and the blackbody temperature corresponding to the summed value was obtained using the Stefan–Boltzmann law and was used as the corrected sky radiant temperature.

*2.4.4 Calculation of $MRT_l$*

We calculated the $MRT_l$ in two different ways. In one of them, the projection of the spherical images was changed to the Lambert cylindrical projection. In this projection, the area on the image is comparable to the solid angle of the surface; namely, the calculation is comparable to an approximation of the human body by a sphere, which is similar to MRT measurement using a black globe thermometer. The $MRT_l$ was calculated as

$$T_{r,Lam} = \sqrt[4]{\frac{\sum_{i=1}^{N_{Lam}}(T_i + 273.15)^4}{N_{Lam}}} - 273.15, \quad (2)$$

where $T_{\mathrm{r,Lam}}$ is the MRT$_{\mathrm{l}}$ calculated using an image in the Lambert cylindrical projection [°C], $N_{\mathrm{Lam}}$ is the number of pixels in the image [-], and $T_i$ is the radiant temperature of the $i$th pixel [°C]. Note that $T_i$ is the radiant temperature, not the surface temperature. Here, we calculate the MRT$_{\mathrm{l}}$ at the measurement position; therefore, the measured radiant temperature, which corresponds to the incident radiation including the influence of surface emissivity, reflected radiation, and atmosphere, is used in equation (2). In order to measure the radiant temperature, the emissivity correction function of the IR camera was not used (emissivity value was set to one).

In the other method, thermal images were projected onto surfaces facing the zenith, the nadir, east, west, south, and north using the orthographic projection. This method is comparable to the approximation of the human body by a cuboid, which is similar to estimating MRT from the plane radiant temperature (PRT) in six opposite directions [19]. In a thermal image with the orthographic projection, the view factor for each surface corresponds to the fraction of pixels in the image, and the longwave PRT was calculated as

$$T_p = \sqrt[4]{\frac{\sum_{i=1}^{N_{ort}}(T_i + 273.15)^4}{N_{ort}}} - 273.15, \quad (3)$$

where $T_p$ is the longwave PRT (PRT$_{\mathrm{l}}$) [°C], and $N_{ort}$ is the number of pixels in the image with the orthographic projection [-]. The MRT$_{\mathrm{l}}$ is given by

$$T_{r,ort} = \sqrt[4]{\sum_{j=1}^{6} w_j (T_{p,j} + 273.15)^4} - 273.15, \quad (4)$$

where $T_{r,ort}$ is the MRT$_{\mathrm{l}}$ calculated using images in the orthographic projection [°C], $w_j$ is the weighting factor for the $j$th direction (projected area factor) [-], and $T_{p,j}$ is the PRT$_{\mathrm{l}}$ for the $j$th direction [°C]. The weighting factors depend on the person's posture [19,34–36].

## 3. Experiments

To validate the developed system, the MRT$_{\mathrm{l}}$ was compared with that obtained using pyrgeometers. We used pyrgeometers as references because they are expected to be the most accurate means of determining MRT$_{\mathrm{l}}$ in outdoor spaces [23,24]. Using pyrgeometers is also suitable for validation in that it can measure longwave radiation from each direction as well as the development system. A globe thermometer cannot separate the effect from shortwave radiation and that from longwave radiation, nor can it measure radiation from each direction separately. Experiments were conducted on the Suzukakedai campus of Tokyo Institute of Technology, Yokohama city, Japan (35.5142°N, 139.4831°E) on October 12, 2020. Our system favors portability over accuracy; thus, there is uncertainty in the treatment of geometric conditions during image synthesis. The accuracy of image synthesis relies on the quality of the feature points, which depends on the color distribution in the image (i.e., the radiant temperature distribution). Therefore, measurements using the developed system and pyrgeometers were conducted at six points with different characteristics (Fig. 4): (1) a space surrounded by buildings on all sides, (2) a space with trees, (3) an open space, (4) a space with vegetated slopes, (5) a space between buildings, and (6) an indoor space. Three rounds of measurements at these six points were conducted beginning at 13:30, 15:30, and 18:00 local standard time (UTC + 9 h). Each round took 40 min. The air temperature at each point was measured using a thermistor sensor with a data logger (T&D TR-72nw-S, T&D Co., Matsumoto, Japan, Table 2) installed in a double pipe with an aspiration fan as a radiation shield. The measurement height was 1 m. The measured air temperature is shown in Table 3.

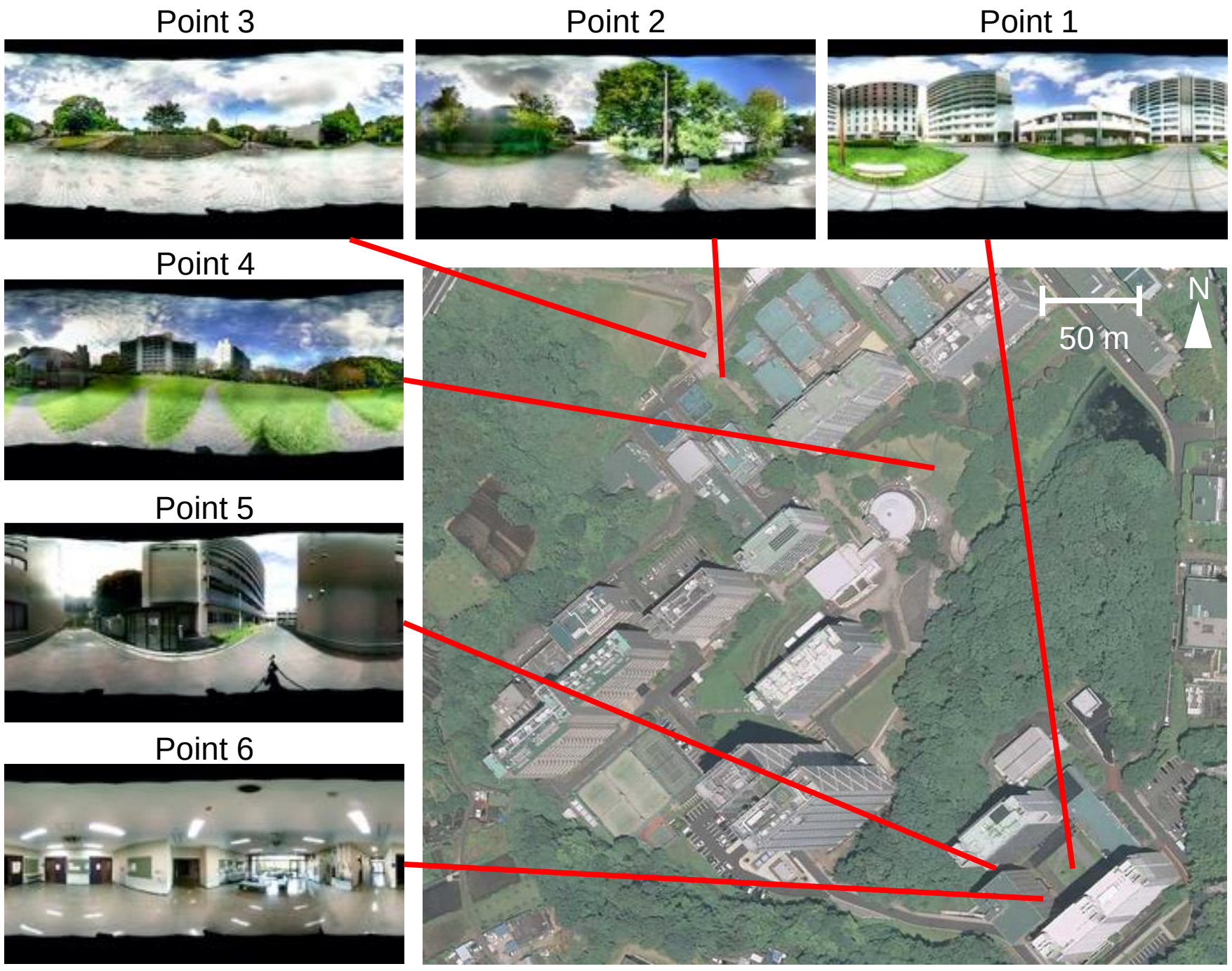

Fig. 4. Photographs of target site. The aerial photograph was taken on June 19, 2019, and was provided by the Geospatial Information Authority of Japan. Spherical visible images of measurement points obtained from the measurements that began at 13:30 are also shown.

Table 2 Specifications of air temperature measurement device.

| | |
|---|---|
| Sensor | Thermistor |
| Measurement range | −25–70 °C |
| Accuracy | ±0.3 °C |
| Measurement resolution | 0.1 °C |
| Response time | 7 min |
| Size | 58 × 78 × 26 mm |
| Weight | 55 g |

Table 3 Air temperature at each time at each point [°C].

| | 13:30 | 15:30 | 18:00 |
|---|---|---|---|
| Point 1 | 24.7 | 25.2 | 24.5 |
| Point 2 | 24.9 | 24.5 | 23.1 |
| Point 3 | 25.2 | 24.4 | 22.6 |
| Point 4 | 25.8 | 24.3 | 22.2 |
| Point 5 | 25.6 | 24.7 | 24.0 |
| Point 6 | 25.3 | 25.6 | 25.0 |

For the developed system, the $MRT_l$ was calculated according to the procedure shown in Section 2. To correct for sky radiant temperature in the developed system, pixels having a radiant temperature value below a threshold were identified as sky pixels. To determine the appropriate threshold value, we manually searched the value where no building pixels are mistakenly identified as sky pixels by conducting identification of sky pixels using different threshold values. The threshold values were determined to be 18 °C, 10 °C, and 10 °C for the measurements at 13:30, 15:30, and 18:00, respectively. The appropriate threshold value is expected to vary with meteorological conditions. Although an objective method is required to identify sky pixels, threshold values were manually determined in this study to identify the sky area. The $PRT_l$ and $MRT_l$ were calculated using equations (2)–(4). For weighting factors in equation (4), we set the same values (1/6) for all directions because the purpose here is validation (comparison with pyrgeometers) rather than the evaluation of $MRT_l$ for actual human bodies, and the directionality can be discussed on the basis of $PRT_l$. Then, any particular posture was not assumed (similar to MRT measurement using a black globe thermometer).

An MR-60 net radiometer (EKO Instruments Co., Ltd., Tokyo, Japan, Table 4) was used. This radiometer measures the shortwave and longwave radiation fluxes from two opposite directions. Our target was longwave radiation; therefore, the signal voltage from the upper and lower pyrgeometers and temperatures from the upper and lower temperature sensors were recorded using a data logger (Thermic Model 2300A, ETO DENKI Co., Tokyo, Japan). The longwave radiation fluxes from six directions (zenith, nadir, east, west, south, and north) were measured by rotating the pyrgeometers. The response time of the pyrgeometer is 18 s. We recorded the measurement value when it stabilized 30–60 s after the measurement began and then rotated pyrgeometers to the next direction.

Table 4 Specifications of the pyrgeometer.

| | |
|---|---|
| Measurement wavelength range | 3–50 μm |
| Field of view | 180° |
| Response time | 18 s |
| Offset due to dome | Maximum of 25 W m$^{-2}$ |
| Size | 60×190×106 mm |
| Weight | 2.8 kg |

According to the manual for the pyrgeometer, its output voltage can be converted to the radiation flux as

$$R_{\text{in}} = \frac{V}{C} + \sigma {T_s}^4, \quad (5)$$

where $R_{\text{in}}$ is the radiation flux to be measured [W m$^{-2}$], $V$ is the output voltage from the pyrgeometer [μV], $C$ is the pyrgeometer sensitivity, which is provided by the manufacturer for each instrument [μV W$^{-1}$ m$^{2}$], $\sigma$ is the Stefan–Boltzmann constant (5.67 × 10$^{-8}$ W m$^{-2}$ K$^{-4}$), and $T_s$ is the temperature from the temperature sensor [K]. Equation (5) does not include the effects of radiation from the dome covering the sensor. Conversion methods considering the dome radiation have been developed [37,38]; they require the dome temperature and constant values related to the dome characteristics. We could not obtain these values; therefore, a simple correction was applied:

$$R_{\text{in}} = \frac{V_{\text{lw}}}{C} + \sigma {T_s}^4 - R_{\text{d}}, \quad (6)$$

where $R_{\text{d}}$ is the radiation from the dome, which we calculated using Planck's law assuming that the emissivity of the dome is zero and one for the measurement wavelength range of the pyrgeometer (3–50 μm) and a longer wavelength, respectively, and the dome temperature was equal to the air temperature. Then, the $PRT_l$ was calculated by converting the radiation flux to the radiant temperature using a relationship between the radiant emittance in the wavelength range of 3–50 μm and the blackbody temperature obtained using Planck's law. The $MRT_l$ was calculated using an equation similar to Eq. (4) with weights similar to the developed system.

## 4. Results

### *4.1 Image synthesis*

Figure 5 shows the synthesized thermal images. It was difficult to automatically identify sufficient feature points for image synthesis in areas where the same pattern appeared repeatedly, such as the tile covering the ground at point 1. However, manual identification was easy in these cases and yielded a synthesized image with no noticeable misalignment (Fig. 5). When automatic identification resulted in errors, the erroneous feature points were removed, and manual identification was applied. For the area with trees at point 2, feature points were well identified automatically. The reason could be that the irregular patterns created by leaves were easy for the machine to decipher, although manual identification was difficult. For the area with uniform radiant temperature (indoor space, point 6), automatic and manual identification of feature points were both difficult. In this case, it was effective to use information on the pan and tilt angles at which each source image was recorded, and the synthesized images had no noticeable misalignment (Fig. 5). The temperature range in the thermal image varies with position and time, so the appropriate temperature range (i.e., color scale) was applied to the images with the same position and time to maximize the nonuniformity of colors within the images for synthesis. In general, the temperature range is higher in outdoor spaces during sunny daytime, and it is easier to identify feature points within the images because of larger temperature differences for the walls and ground.

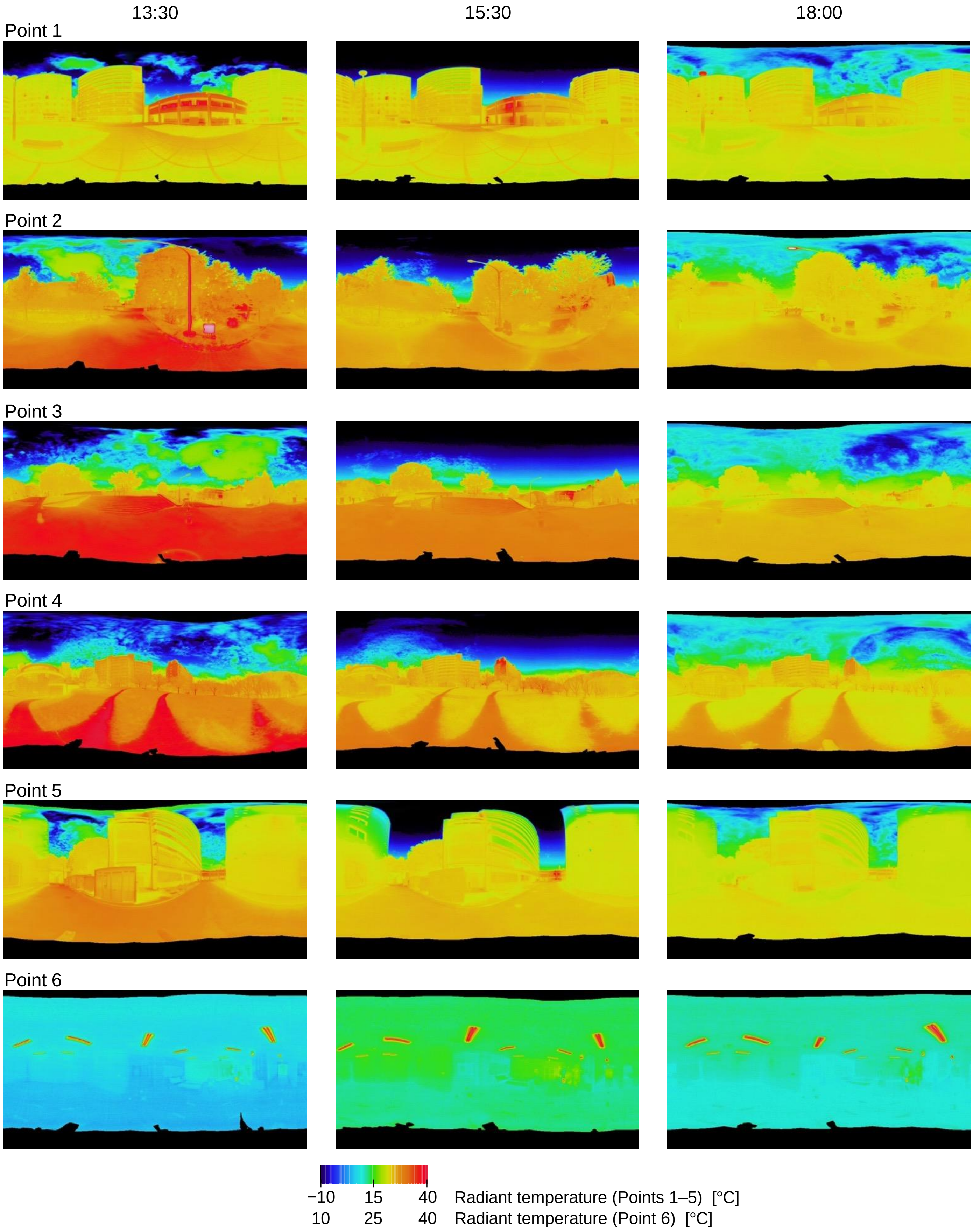
13:30
15:30
18:00
Point 1
Point 2
Point 3
Point 4
Point 5
Point 6
−10 15 40 Radiant temperature (Points 1–5) [°C]
10 25 40 Radiant temperature (Point 6) [°C]

Fig. 5. Spherical thermal images represented in equirectangular projection. The center of the image corresponds to the north direction.

*4.2 $MRT_l$ and $PRT_l$*

Figure 6 shows the $MRT_l$s obtained using the developed system and pyrgeometers at each point and time. The difference in $MRT_l$ between the developed system and pyrgeometers ($\Delta MRT_l$; developed system minus pyrgeometers) was larger in the daytime than in the nighttime, except at point 2. The maximum $\Delta MRT_l$ of −1.6 °C was observed at point 5 at 13:30, but $\Delta MRT_l$ was within 1 °C at all points at 18:00. The $MRT_l$ obtained using the developed system tended to be lower than that obtained using pyrgeometers at 13:30, but not at 15:30 and 18:00. The $MRT_l$ did not vary greatly with projection method. The $MRT_l$ and $PRT_l$ at each measurement point are shown in Fig. 7. The spherical thermal images in the orthographic projection are shown in Fig. 8. The characteristics of each point are as follows.

Point 1 was surrounded by buildings on all sides; the southeast-facing building was low-rise, and the others were 9–12 stories high. The ground and building surfaces were composed of tiles, asphalt, and concrete. Part of the ground surface was vegetated. At 13:30 and 15:30, the southeast-facing building showed a higher temperature than the other surfaces, which were shaded for a long time during the daytime and showed uniform temperature distributions. The vegetated ground surface showed a lower temperature than the impervious ground surfaces. However, the temperature differences between them were relatively small because most of these surfaces were in the shade for a long time. Therefore, the $PRT_l$ shows little directivity because of these spatial characteristics. The spherical thermal images show that the amount of cloud cover varied with time. The difference in $PRT_l$ between the developed system and pyrgeometer ($\Delta PRT_l$; developed system minus pyrgeometer) in the zenith direction was small at all times. $\Delta MRT_l$ was −0.9 °C at 13:30 and was smaller at 15:30 and 18:00.

Point 2 was on a concrete road with trees on both sides. At 13:30, the sunlit road surface showed a relatively high temperature, and the temperature of the sunlit trees was higher than the air temperature. Consequently, the $MRT_l$ was higher than that at point 1. $\Delta PRT_l$ in the zenith direction was −4.5 °C, which was the largest value among all points, times, and directions. At 15:30 and 18:00, $\Delta PRT_l$ in the zenith direction was small. Although $\Delta PRT_l$ in the zenith direction was large at 13:30, $\Delta MRT_l$ at 13:30 was smaller

than those at 15:30 and 18:00.

For point 3, the lower hemisphere was occupied by a tiled road, and the upper hemisphere consisted of sky and trees. The road surface had a high temperature at 13:30, and the sky view factor was larger than that at points 1 and 2, resulting in the largest difference in $PRT_l$ between the zenith and nadir directions among all points and times. The developed system captured this large difference well. The effect of atmospheric radiation on $PRT_l$ in the zenith direction seemed to be large because of the large sky area on the sphere. However, $\Delta PRT_l$ in the zenith direction was small at all times. Although $\Delta PRT_l$ varied slightly with direction, $\Delta MRT_l$ was within 0.5 °C at all times.

For point 4, the lower hemisphere was occupied by a ground surface consisting of a tile path and vegetation. A large area of the upper hemisphere was covered by the sky, similar to point 3. There were buildings in the north-northwest direction and trees in the north-northeast direction. The temperature distribution on the ground was uniform at the other points, but the temperature difference between the tile path and vegetation was significant at point 4. At 13:30, $\Delta PRT_l$ was relatively large in the south direction (−3.2 °C) and varied with direction. This difference in $\Delta PRT_l$ between directions was also observed at 15:30. $\Delta PRT_l$ in the zenith direction was −2.1 °C at 13:30, which was larger than that at 15:30 and 18:00. At 13:30, $\Delta MRT_l$ was −1.5 °C, which was relatively large among points and times. Smaller $\Delta MRT_l$ values were obtained at 15:30 and 18:00.

Point 5 was on an asphalt road with 9–12-story buildings on both sides. The sky view factor was smaller than that at points 1–4. The difference in $PRT_l$ between the zenith and nadir directions obtained using pyrgeometers was approximately one-third to one-fourth that at points 3 and 4 at all times. The developed system overestimated the difference in $PRT_l$ between the zenith and nadir directions at 13:30 because of the underestimation of the $PRT_l$ in the zenith direction. The temperature distributions on the building and ground surfaces were generally uniform, but the upper part of building on the south side, with tile walls, showed a low temperature because of the reflection of atmospheric radiation. $\Delta MRT_l$ was −1.6 °C at 13:30 and decreased with time.

Point 6 was in an indoor space. The surface on the east side was covered with glass. The temperature distributions on the surfaces of the floor, wall, glass, and ceiling were very uniform, except for the lighting, which had high temperatures, at all times. The differences in $\Delta PRT_l$ were all small. The maximum ΔMRT$_l$ at this point was −1.2 °C at 13:30.

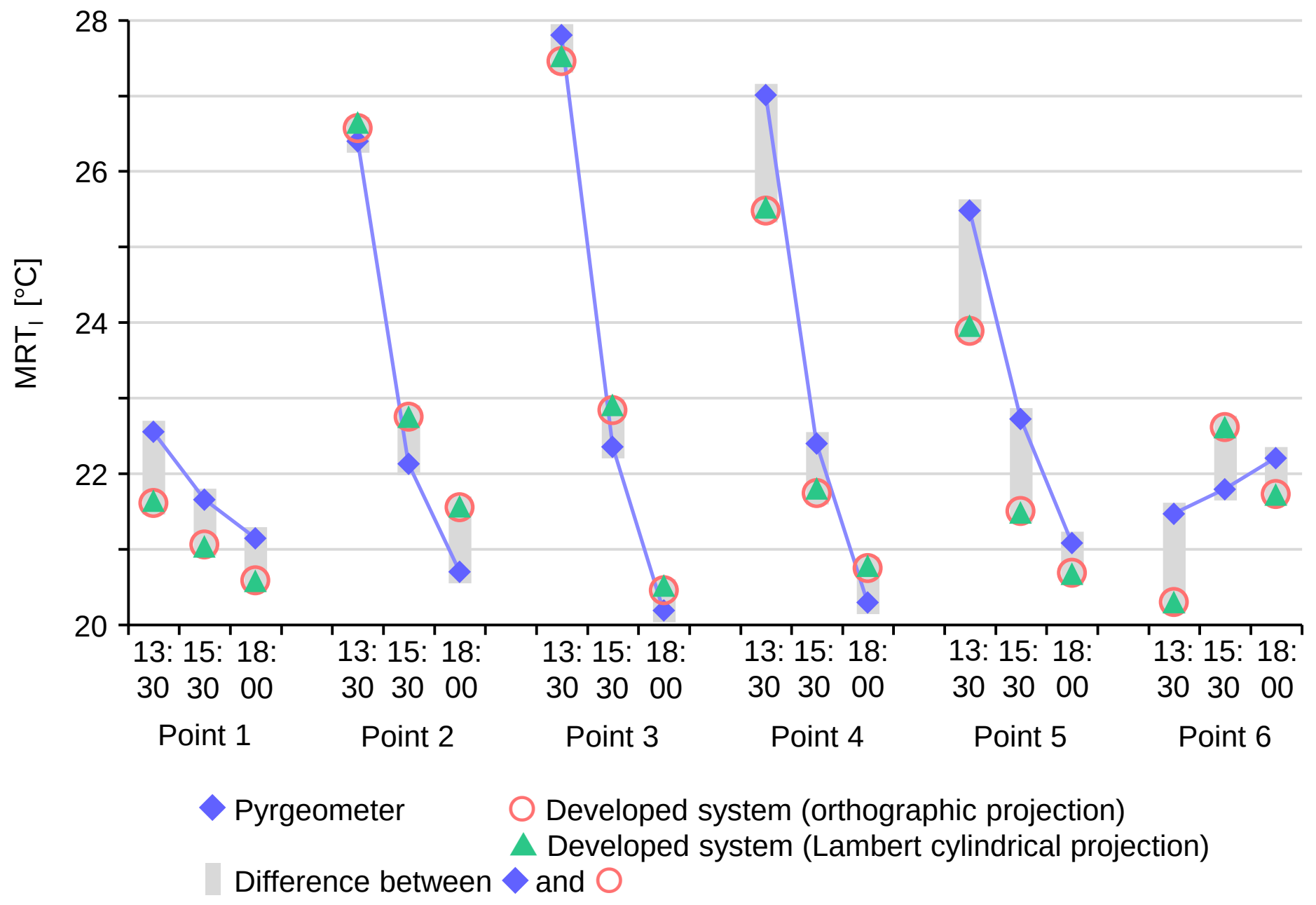

Fig. 6. Comparison of MRT$_l$s obtained by pyrgeometers and developed system.

**Point 1**

| | | 13:30 Pyrgeometer | 13:30 Developed system | 13:30 Δ | 15:30 Pyrgeometer | 15:30 Developed system | 15:30 Δ | 18:00 Pyrgeometer | 18:00 Developed system | 18:00 Δ |
|---|---|---|---|---|---|---|---|---|---|---|
| $PRT_l$ [°C] | Zenith | 19.5 | 18.7 | −0.8 | 18.1 | 17.5 | −0.6 | 19.3 | 18.5 | −0.7 |
| | North | 22.8 | 23.0 | 0.2 | 22.7 | 22.1 | −0.6 | 20.8 | 21 | 0.2 |
| | East | 23.6 | 22.9 | −0.7 | 22.8 | 21.6 | −1.1 | 21.1 | 20.8 | −0.3 |
| | South | 22.6 | 21.3 | −1.3 | 21.7 | 21.6 | −0.0 | 22.1 | 21.1 | −1.0 |
| | West | 22.9 | 21.7 | −1.2 | 21.9 | 21.7 | −0.1 | 22.1 | 21.1 | −1.1 |
| | Nadir | 23.8 | 22.0 | −1.8 | 22.7 | 21.7 | −1.0 | 21.4 | 21 | −0.4 |
| $MRT_l$ [°C] | Ort. | 22.6 | 21.6 | −0.9 | 21.7 | 21.1 | −0.6 | 21.1 | 20.6 | −0.6 |
| | Lam. | 22.6 | 21.6 | −0.9 | 21.7 | 21.0 | −0.6 | 21.1 | 20.6 | −0.6 |

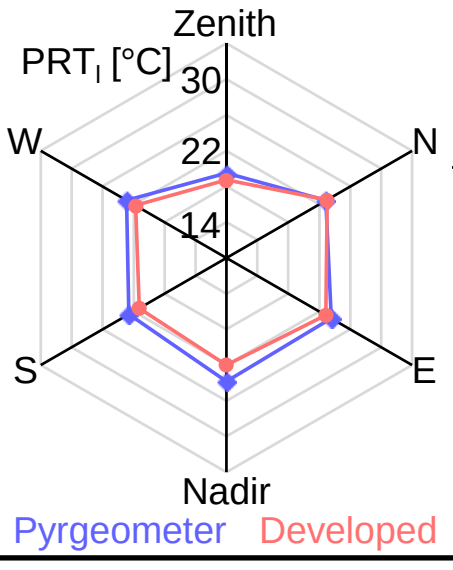

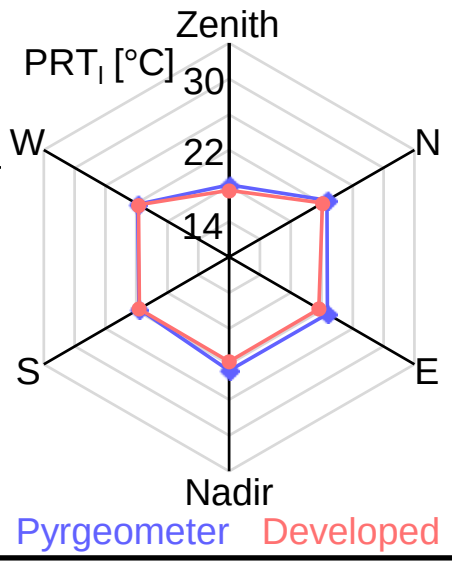
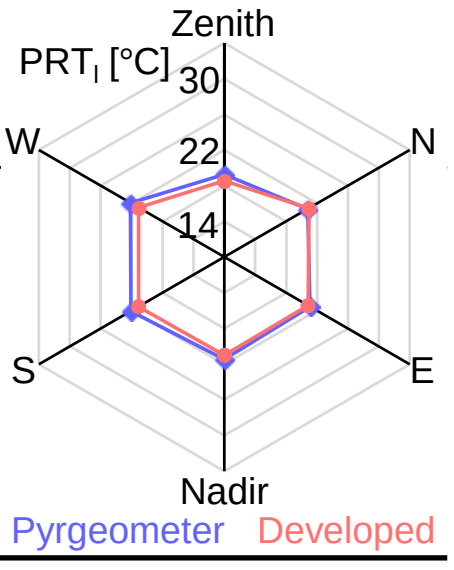

**Point 2**

| | | 13:30 Pyrgeometer | 13:30 Developed system | 13:30 Δ | 15:30 Pyrgeometer | 15:30 Developed system | 15:30 Δ | 18:00 Pyrgeometer | 18:00 Developed system | 18:00 Δ |
|---|---|---|---|---|---|---|---|---|---|---|
| $PRT_l$ [°C] | Zenith | 25.4 | 20.9 | −4.5 | 17.4 | 17.9 | 0.5 | 18.2 | 18.1 | −0.1 |
| | North | 27.0 | 28.5 | 1.5 | 22.1 | 23.3 | 1.2 | 20.8 | 22.4 | 1.6 |
| | East | 26.0 | 27.9 | 1.9 | 23.7 | 24.2 | 0.5 | 21.0 | 22.1 | 1.1 |
| | South | 24.0 | 24.7 | 0.7 | 22.8 | 23.2 | 0.4 | 21.2 | 21.5 | 0.3 |
| | West | 24.7 | 25.5 | 0.8 | 22.5 | 22.5 | 0.0 | 20.9 | 21.4 | 0.5 |
| | Nadir | 31.1 | 31.5 | 0.4 | 24.2 | 25.3 | 1.1 | 22.1 | 23.7 | 1.6 |
| $MRT_l$ [°C] | Ort. | 26.4 | 26.6 | 0.2 | 22.1 | 22.8 | 0.6 | 20.7 | 21.6 | 0.9 |
| | Lam. | 26.4 | 26.6 | 0.2 | 22.1 | 22.8 | 0.6 | 20.7 | 21.6 | 0.9 |

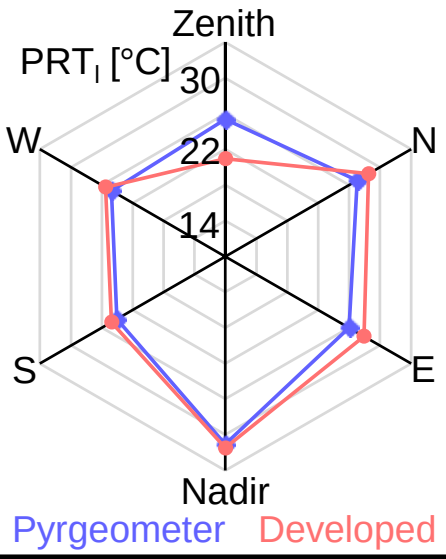
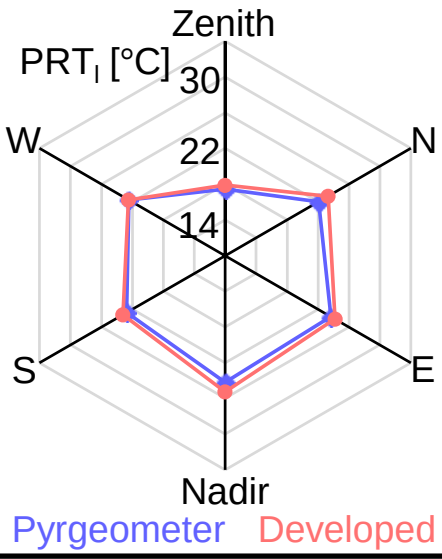
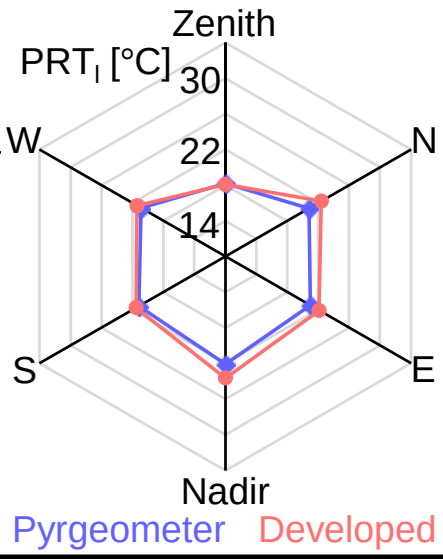

**Point 3**

| | | 13:30 Pyrgeometer | 13:30 Developed system | 13:30 Δ | 15:30 Pyrgeometer | 15:30 Developed system | 15:30 Δ | 18:00 Pyrgeometer | 18:00 Developed system | 18:00 Δ |
|---|---|---|---|---|---|---|---|---|---|---|
| $PRT_l$ [°C] | Zenith | 18.4 | 18.6 | 0.2 | 14.7 | 14.8 | 0.2 | 16.0 | 15.7 | −0.3 |
| | North | 27.7 | 28.4 | 0.7 | 22.3 | 23.4 | 1.1 | 20.2 | 20.9 | 0.7 |
| | East | 28.2 | 27.4 | −0.8 | 21.9 | 22.9 | 1.1 | 20.0 | 20.5 | 0.5 |
| | South | 29.4 | 27.0 | −2.4 | 23.7 | 23.3 | −0.3 | 20.9 | 20.7 | −0.2 |
| | West | 29.3 | 27.7 | −1.6 | 24.0 | 23.8 | −0.2 | 20.8 | 20.9 | 0.2 |
| | Nadir | 33.1 | 34.9 | 1.8 | 27.2 | 28.3 | 1.1 | 23.1 | 23.9 | 0.8 |
| $MRT_l$ [°C] | Ort. | 27.8 | 27.5 | −0.3 | 22.4 | 22.8 | 0.5 | 20.2 | 20.5 | 0.3 |
| | Lam. | 27.8 | 27.5 | −0.3 | 22.4 | 22.9 | 0.6 | 20.2 | 20.5 | 0.3 |

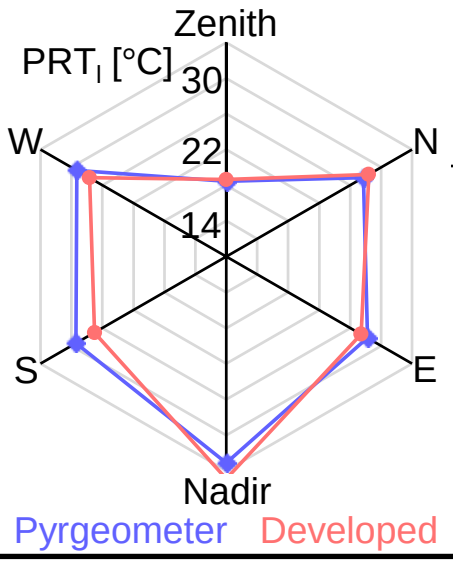
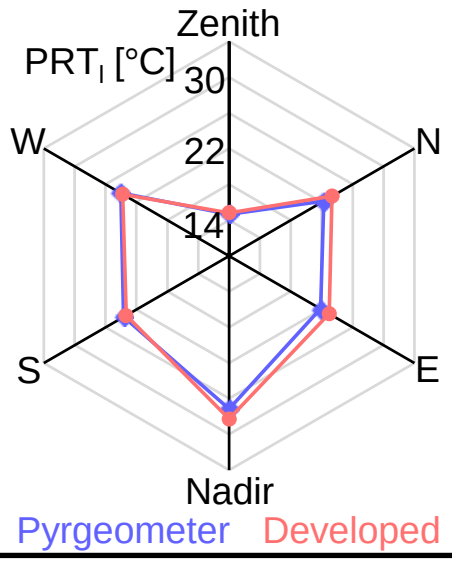
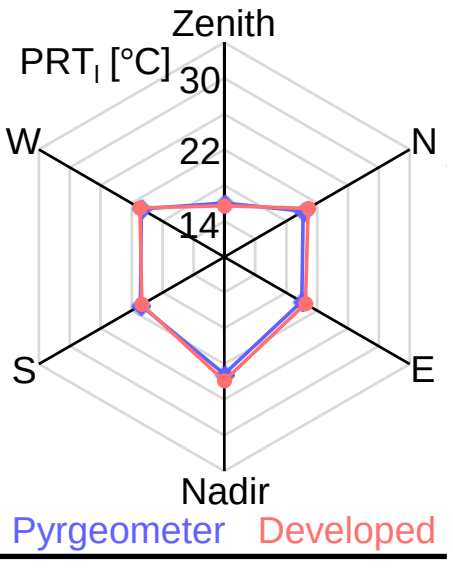

**Point 4**

| | | 13:30 Pyrgeometer | 13:30 Developed system | 13:30 Δ | 15:30 Pyrgeometer | 15:30 Developed system | 15:30 Δ | 18:00 Pyrgeometer | 18:00 Developed system | 18:00 Δ |
|---|---|---|---|---|---|---|---|---|---|---|
| $PRT_l$ [°C] | Zenith | 19.9 | 17.8 | −2.1 | 15.2 | 15.5 | 0.3 | 16.7 | 16.3 | −0.4 |
| | North | 27.0 | 26.3 | −0.7 | 21.7 | 22.7 | 1.0 | 20.3 | 21.6 | 1.2 |
| | East | 25.1 | 24.6 | −0.4 | 20.5 | 21.3 | 0.8 | 19.7 | 20.4 | 0.7 |
| | South | 28.5 | 25.3 | −3.2 | 24.1 | 21.8 | −2.3 | 21.0 | 20.7 | −0.3 |
| | West | 28.1 | 26.1 | −2.0 | 25.6 | 22.6 | −2.9 | 21.0 | 21.2 | 0.2 |
| | Nadir | 33.1 | 32.3 | −0.8 | 26.9 | 26.2 | −0.7 | 22.9 | 24.1 | 1.2 |
| $MRT_l$ [°C] | Ort. | 27.0 | 25.5 | −1.5 | 22.4 | 21.7 | −0.7 | 20.3 | 20.7 | 0.5 |
| | Lam. | 27.0 | 25.5 | −1.5 | 22.4 | 21.8 | −0.6 | 20.3 | 20.8 | 0.5 |

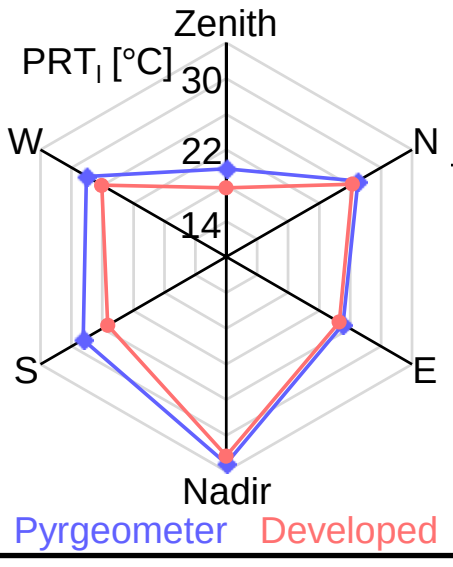
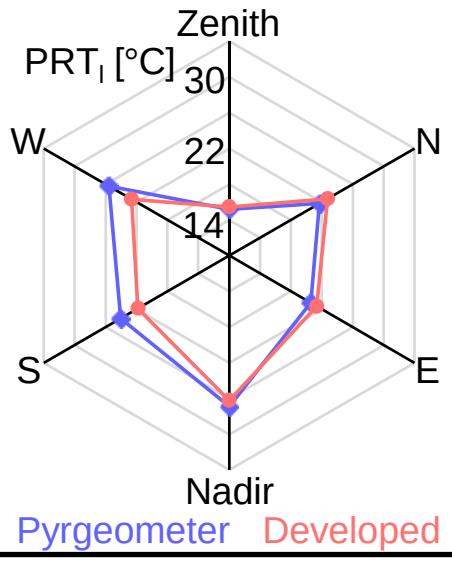
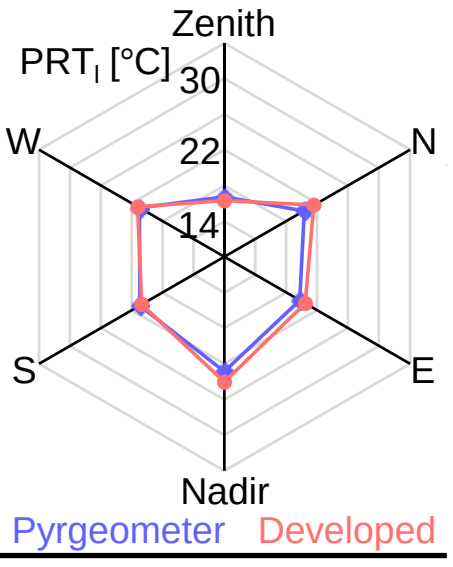

**Point 5**

| | | 13:30 Pyrgeometer | 13:30 Developed system | 13:30 Δ | 15:30 Pyrgeometer | 15:30 Developed system | 15:30 Δ | 18:00 Pyrgeometer | 18:00 Developed system | 18:00 Δ |
|---|---|---|---|---|---|---|---|---|---|---|
| $PRT_l$ [°C] | Zenith | 24.7 | 21.1 | −3.6 | 20.3 | 18.6 | −1.7 | 20.1 | 19.2 | −0.9 |
| | North | 25.2 | 24.4 | −0.7 | 22.9 | 22.2 | −0.7 | 21.3 | 21.3 | 0.0 |
| | East | 23.6 | 23.4 | −0.2 | 22.2 | 21.4 | −0.8 | 20.8 | 20.4 | −0.4 |
| | South | 25.3 | 23.3 | −2.0 | 23.5 | 21.7 | −1.8 | 21.2 | 20.8 | −0.4 |
| | West | 26.5 | 24.6 | −1.9 | 23.3 | 21.7 | −1.5 | 21.1 | 20.8 | −0.3 |
| | Nadir | 27.5 | 26.4 | −1.1 | 24.2 | 23.4 | −0.8 | 22.0 | 21.6 | −0.4 |
| $MRT_l$ [°C] | Ort. | 25.5 | 23.9 | −1.6 | 22.7 | 21.5 | −1.2 | 21.1 | 20.7 | −0.4 |
| | Lam. | 25.5 | 24.0 | −1.5 | 22.7 | 21.5 | −1.2 | 21.1 | 20.7 | −0.4 |

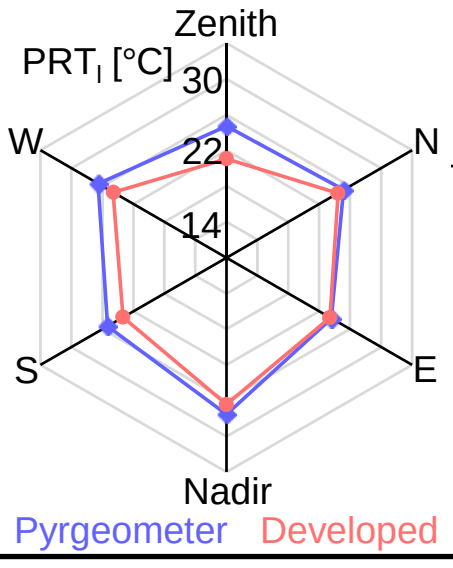
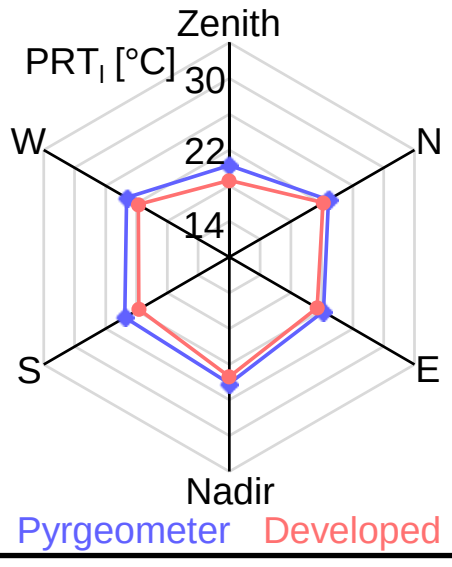
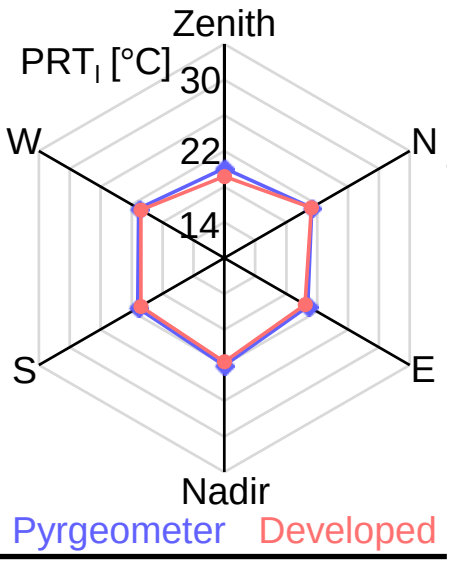

**Point 6**

| | | 13:30 Pyrgeometer | 13:30 Developed system | 13:30 Δ | 15:30 Pyrgeometer | 15:30 Developed system | 15:30 Δ | 18:00 Pyrgeometer | 18:00 Developed system | 18:00 Δ |
|---|---|---|---|---|---|---|---|---|---|---|
| $PRT_l$ [°C] | Zenith | 22.0 | 20.9 | −1.1 | 22.0 | 22.8 | 0.8 | 23.6 | 22.1 | −1.4 |
| | North | 21.6 | 20.2 | −1.3 | 22.2 | 22.5 | 0.4 | 21.9 | 21.6 | −0.3 |
| | East | 21.5 | 20.5 | −1.0 | 22.6 | 22.8 | 0.2 | 21.8 | 21.7 | 0.0 |
| | South | 21.2 | 20.3 | −0.9 | 21.3 | 22.6 | 1.4 | 22.2 | 21.7 | −0.5 |
| | West | 21.3 | 20.3 | −1.0 | 21.4 | 22.6 | 1.2 | 22.4 | 21.8 | −0.6 |
| | Nadir | 21.3 | 19.6 | −1.7 | 21.4 | 22.3 | 0.9 | 21.4 | 21.3 | 0.0 |
| $MRT_l$ [°C] | Ort. | 21.5 | 20.3 | −1.2 | 21.8 | 22.6 | 0.8 | 22.2 | 21.7 | −0.5 |
| | Lam. | 21.5 | 20.3 | −1.2 | 21.8 | 22.6 | 0.8 | 22.2 | 21.7 | −0.5 |

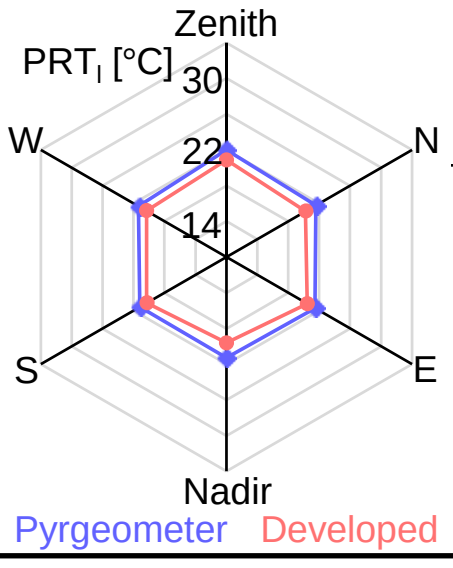
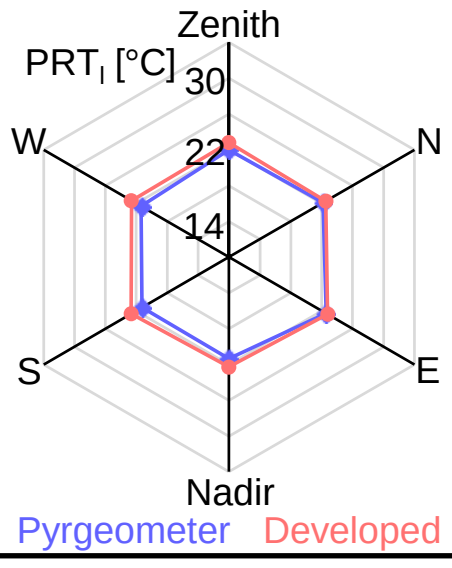
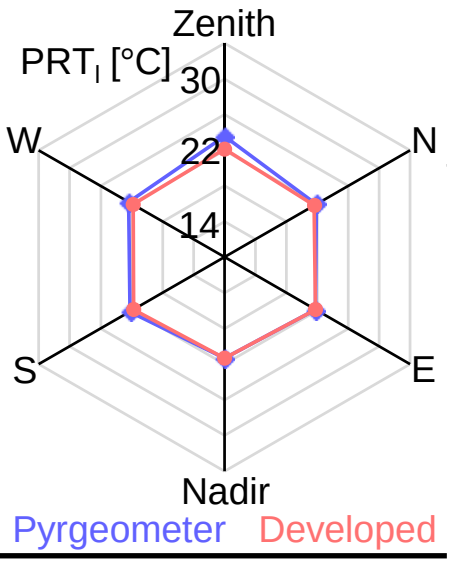

Fig. 7. Differences in $PRT_l$ and $MRT_l$ between pyrgeometers and developed system. The Δ column shows the $\Delta PRT_l$ or $\Delta MRT_l$. For the developed system, $MRT_l$ values derived from thermal images in the orthographic projection (Ort.) and Lambert cylindrical projection (Lam.) are shown in the top and bottom $MRT_l$ rows, respectively. The projection method does not matter for the pyrgeometer, so the two rows contain the same $MRT_l$ value derived from the pyrgeometer.

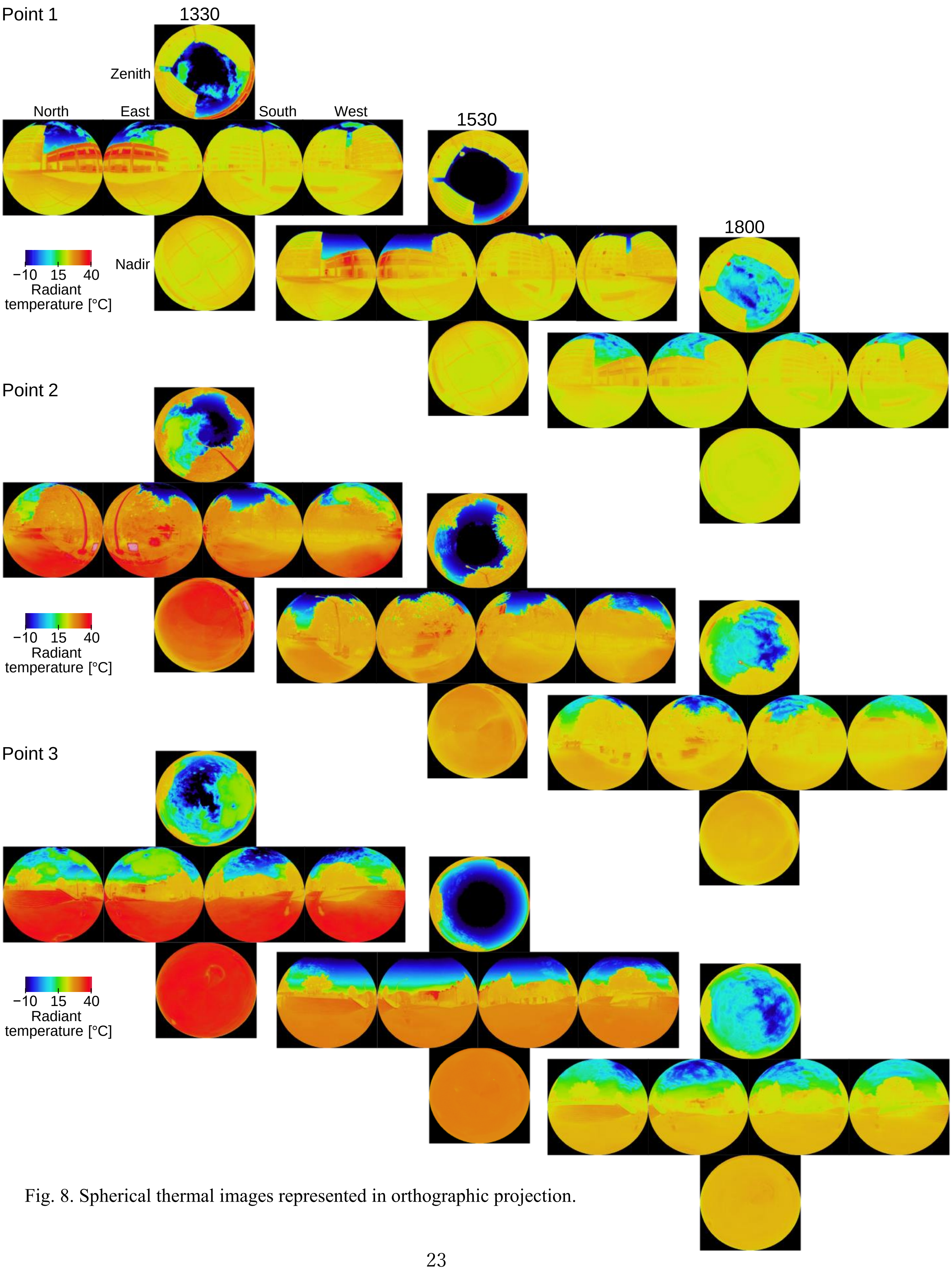

Fig. 8. Spherical thermal images represented in orthographic projection.

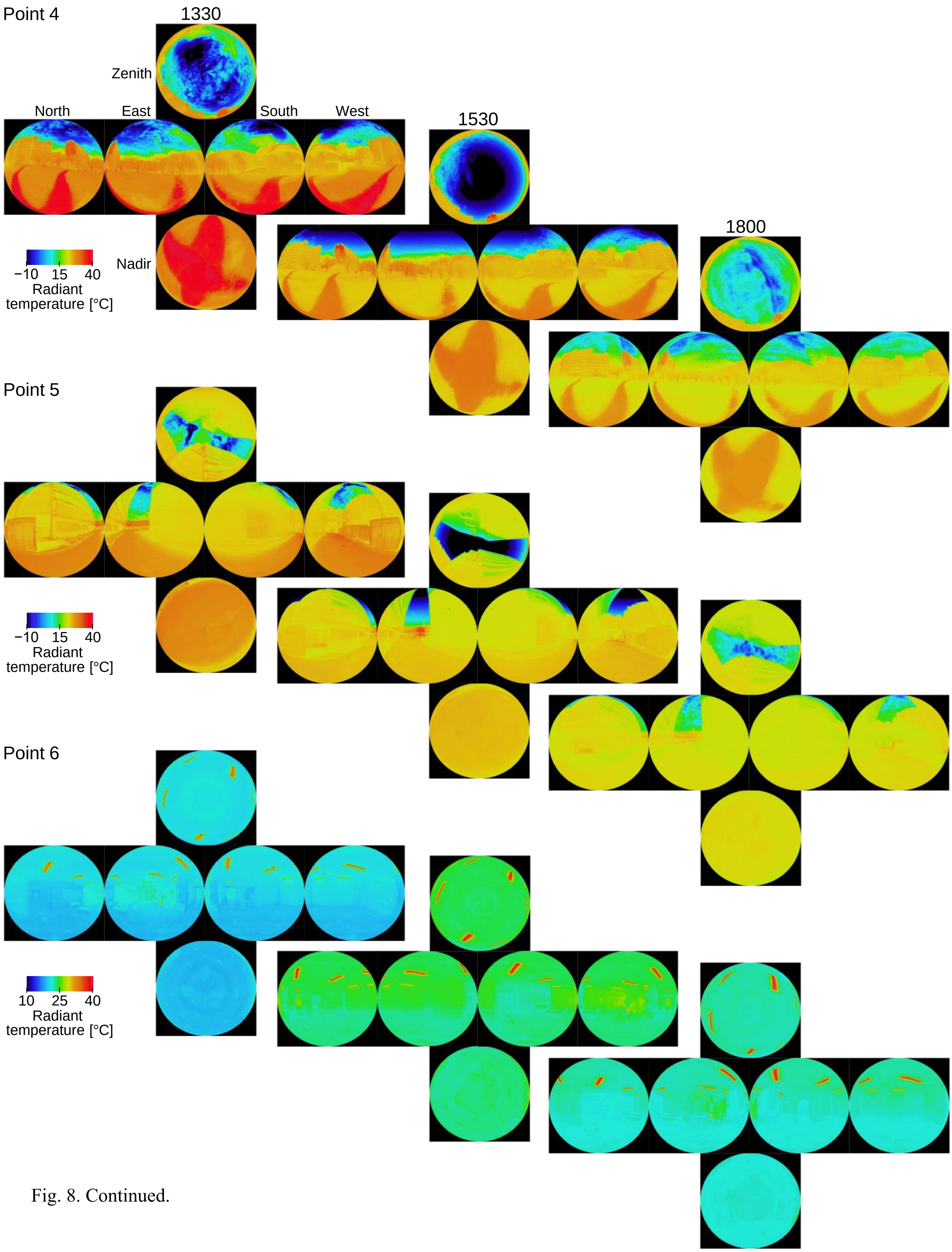

Fig. 8. Continued.

## 5. Discussion

### *5.1 Difference in $MRT_l$ between developed system and pyrgeometers*

For the indoor space (point 6), the correction for sky radiant temperature was not performed, and the temperature distributions on the floor, wall, and ceiling surfaces were small. Therefore, $\Delta MRT_l$ was affected very little by the error in image synthesis, the temperature correction for blind areas, and a slight difference in installation angle (the level and orientation of the instrument) between the developed system and pyrgeometers. $\Delta PRT_l$ was almost constant regardless of direction (Fig. 7); that is, there was a certain difference in temperature between the IR camera and pyrgeometers at each time. This difference resulting from the use of different instruments was estimated to be within approximately 1 °C ($\Delta MRT_l$ was −1.2 °C, 0.8 °C, and −0.5 °C at 13:30, 15:30, and 18:00, respectively). Although the $\Delta MRT_l$ values for the outdoor points were comparable with that for the indoor point, much larger $\Delta PRT_l$ values were observed at the outdoor points.

One difference from the indoor point was that the sky radiant temperature affected the $PRT_l$ of the outdoor spaces. It was sunny with a few clouds at 15:30, and it was cloudy at 18:00. At these times, $\Delta PRT_l$ in the zenith direction was very small at points 1–4 (point 5 was not considered here because the reflection of atmospheric radiation at the building surface seemed to result in a large $\Delta PRT_l$). Therefore, the correction for sky radiant temperature seemed to work well, although an objective method for identifying sky pixels is needed. At 13:30, $\Delta PRT_l$ in the zenith direction was large at points 2 and 4 (4.5 °C and 2.1 °C, respectively). $\Delta PRT_l$ in the nadir direction was less than 1 °C at both points, suggesting that solar radiation might affect the $PRT_l$ in the zenith direction. In fact, the shadow of the measurement instruments clearly appeared in the visible images obtained at 13:30 at points 2 and 4 (Fig. 4). According to the manual for the pyrgeometer, the effect of dome heating due to solar radiation on the measured radiation flux is 25 W m$^{-2}$ at maximum for solar radiation of 1000 W m$^{-2}$. This value corresponds to a radiant temperature of 4.4 °C. The correction term [$R_d$ in Eq. (6)] was approximately 12 W m$^{-2}$ (2.1 °C) in this study. Therefore, the large $\Delta PRT_l$ at point 4 at 13:30 can be attributed to the effect of solar radiation. At point 2, $\Delta PRT_l$ was even larger. According to the thermal image of point 2 in the orthographic projection (Fig. 8), the ratio of sky pixels in the zenith direction was higher than that in the north, east, south, and west directions. However, pyrgeometers showed comparable temperatures in these five directions. Therefore, the difference in installation angle between the developed system and pyrgeometers appeared to contribute to $\Delta PRT_l$, as shown below.

Another difference from the indoor point was the broad temperature distributions on surrounding surfaces and sky, especially at 13:30. Large $\Delta PRT_l$ and differences in $\Delta PRT_l$ between directions were observed at points 2, 3, 4, and 5 at 13:30 and points 4 and 5 at 15:30. For point 5, the atmospheric radiation reflected from the wall of the south building, where the sky view factor was high, seemed to cause the large $\Delta PRT_l$ in the south and west directions. For points 2, 3, and 4, the error in image synthesis and the difference in installation angle appeared to cause the differences in $\Delta PRT_l$ with direction. In the developed system, an error in installation angle is expected to cause an error in the $PRT_l$ but not the $MRT_l$, because the error in the $PRT_l$ in each direction is canceled out when the $MRT_l$ is calculated by summing the $PRT_l$ values. It is difficult to evaluate separately the effects of errors in image synthesis and installation angle on the $PRT_l$ and $MRT_l$. The effect of errors in installation angle was evaluated as follows.

The variation in the radiation flux on the surface of a cube with rotation of the cube was calculated in a virtual space with the same surface temperature distribution as the synthesized thermal image (after corrections for sky radiant temperature and blind area). The calculation was performed for a right-hand system with the *x* axis pointing north, and the variations in the $PRT_l$ and $MRT_l$ with rotation around the *x*, *y*, and *z* axes were obtained. The results are presented in Fig. 9. The variation in $PRT_l$ was large at points 2, 3, and 4, where the variation in surface temperature in the space was large. At points 3 and 4, with a large sky area in the upper hemisphere, the variation in $PRT_l$ in the zenith direction was small. By contrast, the variation in $PRT_l$ in the zenith direction was larger for point 2, where trees occupied areas with smaller zenith angles. However, the variation was too small to fully explain the large $\Delta PRT_l$ at point 2 at 13:30. For lateral directions, even a small rotation can cause the $PRT_l$ to vary by more than 0.5 °C at points 2, 3, and 4. As expected, the $MRT_l$ was affected very little by rotation. However, rotation affected the $MRT_l$ obtained using pyrgeometers, which has upper and lower sensors, because the three measurements are independent of each other.

In conclusion, the maximum difference in $MRT_l$ due to the instruments was approximately 1 °C. $\Delta MRT_l$ was within 1 °C in most cases and was 1.6 °C at maximum. The errors of pyrgeometers include errors due to solar radiation, installation angle, and reflected atmospheric radiation, although it is considered a more accurate method of measuring the $MRT_l$, as described in the Introduction. The difference of 1.6 °C is a sufficient indication of the accuracy of the developed system. The installation angle error of the developed system, which is a few degrees, can affect the $PRT_l$ by more than 0.5 °C, but does not affect the $MRT_l$.

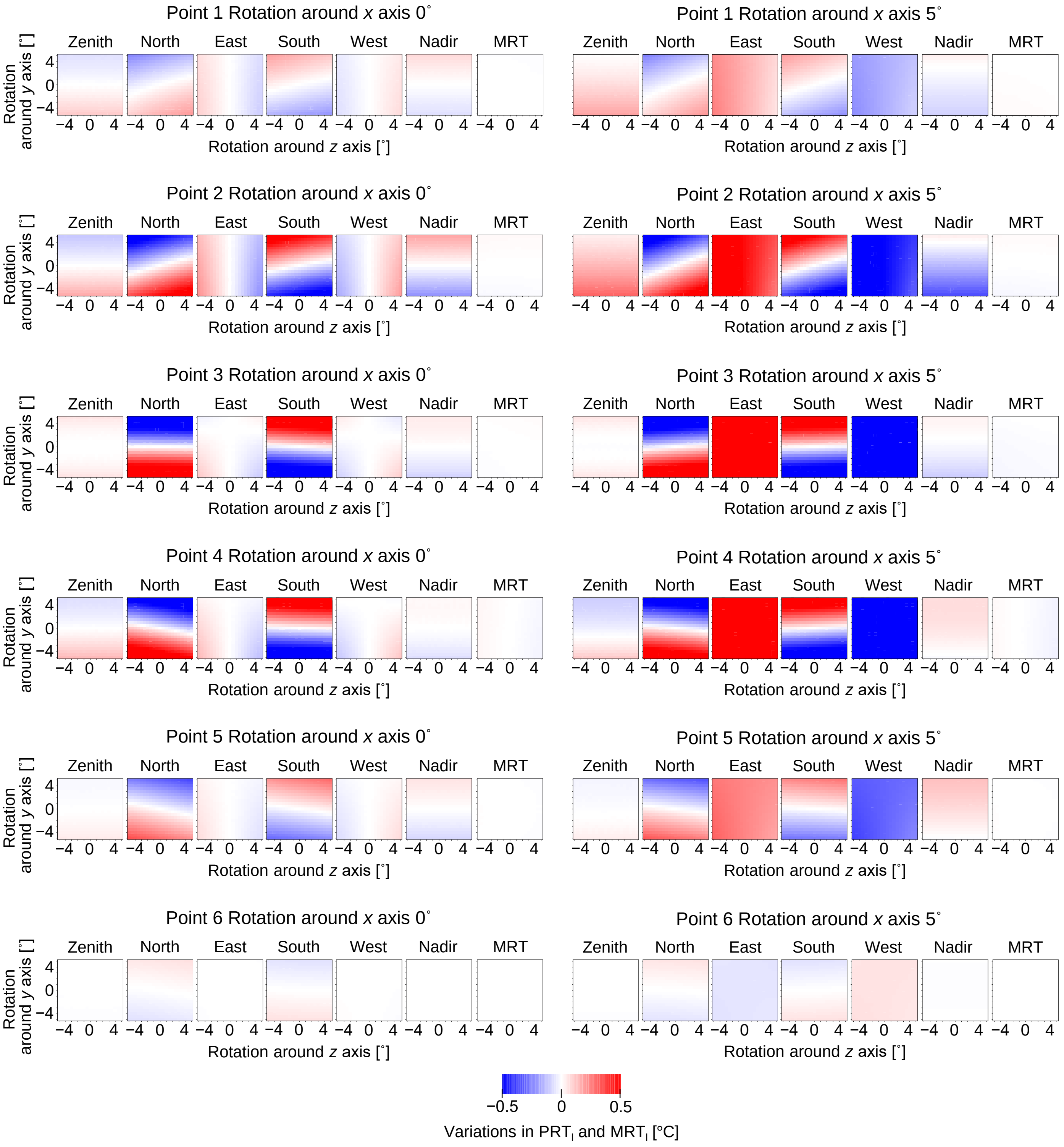

Fig. 9. Variations in $PRT_l$ and $MRT_l$ with rotation of a virtual cube to calculate $PRT_l$ and $MRT_l$. Results of six surfaces and $MRT_l$ for rotation around $x$ axis by 0° and 5° are shown. The color indicates $T_{\theta,\varphi,\phi} - T_{0,0,0}$, where $T_{\theta,\varphi,\phi}$ is the $PRT_l$ or $MRT_l$ at rotation angles around the $x$, $y$, and $z$ axes of $\theta$, $\varphi$, and $\phi$, respectively.

*5.2 Comparison with previous studies*

Regarding the measurement accuracy of a spherical thermography system, Asano et al. [27] did not compare their results with those obtained using other methods. Paz y Miño et al. [28] compared their MRT with that obtained using a globe thermometer and reported that the difference was always within 2 °C for indoor and outdoor measurements in summer and winter. For other MRT measurement systems, Revel et al. [32] tested a thermopile scanning system in an indoor space and showed that the difference in MRT between the system and a globe thermometer was 0.0 ± 0.5 °C and 0.4 ± 2.0 °C under conditions where solar radiation had small and large effects, respectively (in the latter case, correction for shortwave radiation was applied). Lee and Jo [29] measured the MRT by scanning an IR camera in indoor space, but they compared the obtained surface temperature only with that measured using a contact thermometer at several points. Alfano et al. [20] compared MRT values obtained by different methods in an indoor space and showed that the difference in MRT between an IR camera and a net radiometer was 0.7 °C. We tested the developed system in outdoor and indoor environments, including conditions with solar radiation and reflected atmospheric radiation. $\Delta MRT_l$ was smaller than 1 °C in most cases and was 1.6 °C at maximum, which is effectively comparable with the results of these previous studies.

*5.3 Ease of use of developed system*

The pioneering spherical thermography system developed by Asano et al. [27] was very large and heavy and used a custom-made rotation platform. It was labor-intensive to move and assemble the system. The scanning system developed by Lee and Jo [29] is similar in appearance to that of Asano et al. [27], although details regarding portability and ease of use were not reported. Paz y Miño et al. [28] employed a commercially available rotation platform that weighs 3 kg in their scanning system. Their system seemed to offer greatly improved portability; however, the platform was not inexpensive. The system we developed weighs only 2.7 kg and can easily be carried. The time required for preparation and measurement at one point was 5 min, and the measurements at six points reported in this study were completed within 40 min. In an actual urban district in the center of Tokyo, we conducted measurements using this system every 2 h (from 10:00 to 20:00) at 15 points within 1.5 h (the results are shown at the following URL: https://360.goterest.com/sphere/marunouchi-2nd-street?scene=5dd15fa2ffd0e1bd21a5cddc). The features of being portable and easy to use are particularly useful when the target area is large with complex spatial geometry, such as in actual urban spaces, rather than in simple indoor spaces where measurement in one or a few locations is sufficient. The battery lifetimes of the camera and rotation platform are 4 h and 8–14 h, respectively, so it is possible to continuously conduct measurements at more than 30 points in an actual urban district. The camera and rotation platform are commercially available and inexpensive, and the system is easily controlled by an app.

In outdoor spaces, it is expected that the surrounding conditions may change from time to time, depending on the users of the space, passersby, automobiles, etc. The spherical thermal image obtained by the developed system is not a snapshot, i.e., moving objects may be included in more than one source image. However, this caution also applies to other instruments and methods. The advantage of this system is that it is easy and quick to use, so it can be used to observe thermal radiation environments under such a variety of conditions.

*5.4 Limitations*

Regarding the image synthesis, feature points were well identified automatically for most surfaces. However, manual identification was required for some surfaces, such as tiles with uniform patterns. The advantage of the developed system is the ease of measurement. It is assumed that many images will be taken in the practical use as shown in the previous section. Therefore, even though identifying feature points manually itself is not difficult, it will be possible to synthesize with less manual processing. In the present study, an image stitching software was used for image synthesis. In the future, we will develop a synthesis method specific to thermal images so that spherical images can be synthesized automatically.

One of the important points to note when using IR cameras is their measurement wavelength range. The atmospheric windows are commonly used (8–14 μm for the IR camera in the developed system). Materials with a large difference in emissivity between the measurement wavelength range and other regions are error factors in the $MRT_l$ measurement using IR cameras. In the present study, a correction was conducted for the sky radiant temperature. Some other materials, such as glass with low emissivity at around 10 µm, require caution. If the area of glass is small, the impact on $MRT_l$ measurement is expected to be small as $\Delta PRT_l$ and $\Delta MRT_l$ were small at points 1 and 6 (small spatial distribution of radiant temperature due to indoor space also contributed to this result for point 1). However, caution should be exercised when applying the developed system to $MRT_l$ measurement in spaces surrounded by large windows. If characteristics of the material (directional and spectral emissivity) and the three-dimensional geometry of the target space are known, it is possible to correct the effect of spectral emissivity using the spherical thermal image as Hoyano and Wakui [39]. Whether to incorporate such correction into the developed system in the future should be determined considering the balance between the difficulty of correction and the system's feature of ease of use.

Although thermography can only evaluate longwave radiation, the advantage is that it can visualize radiant temperature distribution and evaluate the impact of each surface on MRT. Shortwave radiation, for which it is important whether the surface is in sun or shade, is not that difficult to evaluate using models, as in the method of [30,31]. Nevertheless, it is worthwhile to be able to measure shortwave MRT together with $MRT_l$. We plan to consider shortwave radiation in our system, but the following issues must be addressed. Shortwave radiation has a very broad brightness distribution because of direct solar radiation and sky-diffuse radiation around the sun, making it difficult to capture with a single camera. Paz y Miño et al. [28] used an estimation model of the sky radiation distribution instead of measured sky data to calculate the MRT under a clear sky and direct solar radiation. It is necessary to use cameras that can handle a wide range of radiance, and a separate estimate is needed for the direct component, which has the highest energy density. For shortwave radiation, it is important to consider not only the visible region but also the near-infrared region; however, the most common cameras are visible imaging cameras. Approximately 50% of the energy of solar radiation falls in the near-infrared region, which cannot be ignored. It is desirable to estimate the radiance in the near-infrared region from visible images, but simple estimation from the visible region is not possible because, for example, vegetation has very high reflectance in the near-infrared region. As a material with similar characteristics, heat-insulating pavement (cool pavement) using paint with high reflectance in the near-infrared region has become popular in recent years as a countermeasure against high-temperature road surfaces. Paz y Miño et al. [28] considered only the visible region and slightly beyond it (380–780 nm). A camera capable of measuring near-infrared radiation up to a wavelength of approximately 3 μm is effective.

## 6. Conclusions

We developed a portable spherical thermography system consisting of commercially available instruments and image processing for panorama synthesis. The system weighs only 2.7 kg and can be easily assembled and controlled. The system records 24 source images within 2 min. A spherical thermal image is generated by synthesizing the source images according to feature points. Feature points can be identified automatically for areas with irregular patterns, such as tree leaves. Automatic identification is difficult for areas with regular patterns, such as road tiles, and uniform radiant temperature. In these cases, manual identification and information on the pan and tilt angles at which each source image were recorded are effective for generating images with no noticeable misalignment. To validate the developed system, measurements using the system were conducted in outdoor and indoor environments with various radiant temperature distributions, and the obtained $PRT_l$ and $MRT_l$ with orthographic projection of spherical thermal images were compared with those obtained using pyrgeometers for six directions. According to the indoor measurements, the difference in $MRT_l$ between the developed system and pyrgeometers due to the performance of the sensors themselves was estimated to be approximately 1 °C. The difference in $MRT_l$ was less than 1 °C in most cases, including outdoor measurements. The effects of solar radiation on pyrgeometers, the error in installation angle of pyrgeometers, and reflected atmospheric radiation incident on the camera affected the $MRT_l$ in some cases; however, the maximum difference was 1.6 °C. Although our system is small and simple, the difference in $MRT_l$ between the system and pyrgeometers was effectively comparable with those of previous systems. A method to automatically synthesize a spherical image from the recorded source images will be developed in the future. We also plan to improve the system so that it can take into account shortwave radiation.

## CRediT authorship contribution statement

**Takashi Asawa:** Conceptualization, Data curation, Funding acquisition, Project administration, Resources, Supervision, Writing - review & editing. **Haruki Oshio:** Formal analysis, Software, Validation, Visualization, Writing - original draft, Writing - review & editing. **Kazuki Tanaka:** Formal analysis, Investigation, Methodology, Software, Writing - review & editing.

**Declaration of competing interest**

The authors declare no competing interests.

**References**

[1] D.A. McIntyre, I.D. Griffiths, Subjective response to radiant and convective environments, Environ. Res. 5 (1972) 471–482. https://doi.org/10.1016/0013-9351(72)90048-5.

[2] H. Guo, D. Aviv, M. Loyola, E. Teitelbaum, N. Houchois, F. Meggers, On the understanding of the mean radiant temperature within both the indoor and outdoor environment, a critical review, Renew. Sust. Energ. Rev. 117 (2020) 109207. https://doi.org/10.1016/j.rser.2019.06.014.

[3] K. Rhee, K.W. Kim, A 50 year review of basic and applied research in radiant heating and cooling systems for the built environment, Build. Environ. 91 (2015), 166–190. https://doi.org/10.1016/j.buildenv.2015.03.040.

[4] H. Kitagawa, T. Asawa, T. Kubota, A.R. Trihamdani, K. Sakurada, H. Mori, Optimization of window design for ventilative cooling with radiant floor cooling systems in the hot and humid climate of Indonesia, Build. Environ. 188 (2021) 107483. https://doi.org/10.1016/j.buildenv.2020.107483.

[5] H. Mayer, P. Höppe, Thermal comfort of man in different urban environments, Theor. Appl. Climatol. 38 (1987) 43–49. https://doi.org/10.1007/BF00866252.

[6] M. Nikolopoulou, S. Lykoudis, Use of outdoor spaces and microclimate in a Mediterranean urban area, Build. Environ. 42 (2007) 3691–3707. https://doi.org/10.1016/j.buildenv.2006.09.008.

[7] L. Chen, E. Ng, Outdoor thermal comfort and outdoor activities: A review of research in the past decade, Cities 29 (2012) 118–125. https://doi.org/10.1016/j.cities.2011.08.006.

[8] H. Andrade, M.J. Alcoforado, Microclimatic variation of thermal comfort in a district of Lisbon (Telheiras) at night, Theor. Appl. Climatol. 92 (2008) 225–237. https://doi.org/10.1007/s00704-007-0321-5.

[9] ASHRAE Handbook—Fundamentals 2001 (SI Edition), American Society of Heating, Refrigerating, and Air-Conditioning Engineers (2001). ISBN: 1883413885.

[10] F. Lindberg, B. Holmer, S. Thorsson, SOLWEIG 1.0 – Modelling spatial variations of 3D radiant fluxes and mean radiant temperature in complex urban settings, Int. J. Biometeorol. 52 (2008) 697–713.

https://doi.org/10.1007/s00484-008-0162-7.

[11] M. Taleghani, D.J. Sailor, M. Tenpierik, A. Dobbelsteen, Thermal assessment of heat mitigation strategies: The case of Portland State University, Oregon, USA, Build. Environ. 73 (2014) 138–150. https://doi.org/10.1016/j.buildenv.2013.12.006.

[12] J. Huang, J.G. Cedeño-Laurent, J.D. Spengler, CityComfort+: A simulation-based method for predicting mean radiant temperature in dense urban areas, Build. Environ. 80 (2014) 84–95. https://doi.org/10.1016/j.buildenv.2014.05.019.

[13] W. Klemm, B.G. Heusinkveld, S. Lenzholzer, B. Hove, Street greenery and its physical and psychological impact on thermal comfort, Landsc. Urban Plan. 138 (2015) 87–98. https://doi.org/10.1016/j.landurbplan.2015.02.009.

[14] C.Y. Park, D.K. Lee, E.S. Krayenhoff, H.K. Heo, S. Ahn, T. Asawa, A. Murakami, H.G. Kim, A multilayer mean radiant temperature model for pedestrians in a street canyon with trees, Build. Environ. 141 (2018) 298–309. https://doi.org/10.1016/j.buildenv.2018.05.058.

[15] S. Thorsson, J. Rocklöv, J. Konarska, F. Lindberg, B. Holmer, B. Dousset, D. Rayner, Mean radiant temperature – A predictor of heat related mortality, Urban Clim. 10 (2014) 332–345. http://dx.doi.org/10.1016/j.uclim.2014.01.004.

[16] F. Lindberg, C.S.B. Grimmond, Nature of vegetation and building morphology characteristics across a city: Influence on shadow patterns and mean radiant temperatures in London, Urban Ecosyst. 14 (2011) 617–634. https://doi.org/10.1007/s11252-011-0184-5.

[17] N. Kántor, C.V. Gál, Á. Gulyás, J. Unger, The impact of façade orientation and woody vegetation on summertime heat stress patterns in a central European square: Comparison of radiation measurements and simulations, Adv. Meteorol. 2018 (2018) 2650642. https://doi.org/10.1155/2018/2650642.

[18] N. Kántor, L. Chen, C.V. Gál, Human-biometeorological significance of shading in urban public spaces—Summertime measurements in Pécs, Hungary, Landsc. Urban Plan. 170 (2018) 241–255. https://doi.org/10.1016/j.landurbplan.2017.09.030.

[19] ISO 7726, Ergonomics of the Thermal Environment – Instruments for Measuring Physical Quantities, International Standardization Organization, Geneva, 1998.

[20] F.R.D. Alfano, M. Dell'Isola, B.I. Palella, G. Riccio, A. Russi, On the measurement of the mean radiant

temperature and its influence on the indoor thermal environment assessment, Build. Environ. 63 (2013) 79–88. http://dx.doi.org/10.1016/j.buildenv.2013.01.026.

[21] F.R.D. Alfano, M. Dell'Isola, G. Ficco, B.I. Palella, G. Riccio, On the measurement of the mean radiant temperature by means of globes: An experimental investigation under black enclosure conditions, Build. Environ. 193 (2021) 107655. https://doi.org/10.1016/j.buildenv.2021.107655.

[22] H. Guo, E. Teitelbaum, N. Houchois, M. Bozlar, F. Meggers, Revisiting the use of globe thermometers to estimate radiant temperature in studies of heating and ventilation, Energy Build. 180 (2018) 83–94. https://doi.org/10.1016/j.enbuild.2018.08.029.

[23] S. Thorsson, F. Lindberg, I. Eliasson, B. Holmer, Different methods for estimating the mean radiant temperature in an outdoor urban setting, Int. J. Climatol. 27 (2007) 1983–1993. https://doi.org/10.1002/joc.1537.

[24] E. Johansson, S. Thorsson, R. Emmanuel, E. Krüger, Instruments and methods in outdoor thermal comfort studies – The need for standardization, Urban Clim. 10 (2014) 346–366. http://dx.doi.org/10.1016/j.uclim.2013.12.002.

[25] H. Guo, M. Ferrara, J. Coleman, M. Loyola, F. Meggers, Simulation and measurement of air temperatures and mean radiant temperatures in a radiantly heated indoor space, Energy 193 (2020) 116369. https://doi.org/10.1016/j.energy.2019.116369.

[26] M.F. Özbey, C. Turhan, A comprehensive comparison and accuracy of different methods to obtain mean radiant temperature in indoor environment, Thermal Sci. Eng. Prog. 31 (2022) 101295. https://doi.org/10.1016/j.tsep.2022.101295.

[27] K. Asano, A. Hoyano, T. Matsunaga, Development of an urban thermal environment measurement system using a new spherical thermography technique, in: Proceedings of SPIE 2744, Infrared Technology and Applications XXII, June 1996, 620–631. https://doi.org/10.1117/12.243504.

[28] J.A. Paz y Miño, C. Lawrence, B. Beckers, Visual metering of the urban radiative environment through 4π imagery, Infrared Phys. Technol. 110 (2020) 103463. https://doi.org/10.1016/j.infrared.2020.103463.

[29] D.-S. Lee, J.-H. Jo, Pan–tilt IR scanning method for the remote measurement of mean radiant temperatures at multi-location in buildings, Remote Sens. 13 (2021) 2158. https://doi.org/10.3390/rs13112158.

[30] D.-S. Lee, J.-H. Jo, Application of simple sky and building models for the evaluation of solar irradiance distribution at indoor locations in buildings, Build. Environ. 197 (2021) 107840. https://doi.org/10.1016/j.buildenv.2021.107840.

[31] D.-S. Lee, J.-H. Jo, Application of IR camera and pyranometer for estimation of longwave and shortwave mean radiant temperatures at multiple locations, Build. Environ. 207 (2022) 108423. https://doi.org/10.1016/j.buildenv.2021.108423.

[32] G.M. Revel, M. Arnesano, F. Pietroni, Development and validation of a low-cost infrared measurement system for real-time monitoring of indoor thermal comfort, Meas. Sci. Technol. 25 (2014) 085101. https://doi.org/10.1088/0957-0233/25/8/085101.

[33] H. Guo, E. Teitelbaum, J. Read, F. Meggers, Mapping comfort with the SMART (Spherical Motion Average Radiant Temperature) sensor, in: Proceedings of Building Simulation 2017, the 15th International Conference of IBPSA, San Francisco CA, USA, Aug. 2017, 7–9.

[34] P.O. Fanger, Thermal Comfort, Danish Technical Press, Copenhagen, 1970.

[35] C. Marino, A. Nucara, G. Peri, M. Pietrafesa, G. Rizzo, A generalized model of human body radiative heat exchanges for optimal design of indoor thermal comfort conditions, Sol. Energy 176 (2018) 556–571. https://doi.org/10.1016/j.solener.2018.10.052.

[36] M.H. Vorre, R.L. Jensen, J. Le Dréau, Radiation exchange between persons and surfaces for building energy simulations, Energy Build. 101 (2015) 110–121. http://dx.doi.org/10.1016/j.enbuild.2015.05.005.

[37] C.W. Fairall, P.O.G. Persson, E.F. Bradley, R.E. Payne, S.P. Anderson, A new look at calibration and use of Eppley precision infrared radiometers. Part I: theory and application, J. Atmos. Ocean. Technol. 15 (1998) 1229–1242. https://doi.org/10.1175/1520-0426(1998)015<1229:ANLACA>2.0.CO;2.

[38] Q. Ji, S. Tsay, On the dome effect of Eppley pyrgeometers and pyranometers, Geophys. Res. Lett. 27 (2000) 972–974. https://doi.org/10.1029/1999GL011093.

[39] A. Hoyano, T. Wakui, Generation of surface temperature image derived from spherical thermograph in urban environment, IEEJ T. Fund. Mater. 128 (2008) 158–163. https://doi.org/10.1541/ieejfms.128.158.